\documentclass[letterpaper,twocolumn,10pt]{article}
\usepackage{usenix-2020-09}
\usepackage[available]{usenixbadges}

\usepackage{tikz}
\usepackage{amsmath}

\usepackage{filecontents}

\usepackage{tikz}
\usepackage{amsmath}
\usepackage{amsfonts}
\usepackage{mathtools}
\usepackage{amssymb}
\usepackage{amsthm}
\usepackage{bm}
\usepackage{bbm} %
\usepackage[english]{babel} %
\usepackage[most]{tcolorbox}

\def\eqref#1{equation~\ref{#1}}

\def\1{\bm{1}}

\DeclareMathAlphabet{\mathsfit}{\encodingdefault}{\sfdefault}{m}{sl}
\SetMathAlphabet{\mathsfit}{bold}{\encodingdefault}{\sfdefault}{bx}{n}

\newcommand{\abs}[1]{ \left\lvert#1\right\rvert} %

\theoremstyle{plain}
\newtheorem{theorem}{Theorem}[section]

\newtheorem{definition}[theorem]{Definition}

\usetikzlibrary{patterns}

\pgfdeclarepatternformonly{dense north west lines}{\pgfqpoint{-1pt}{-1pt}}{\pgfqpoint{2pt}{2pt}}{\pgfqpoint{1.5pt}{1.5pt}}%
{
  \pgfsetlinewidth{0.4pt}
  \pgfpathmoveto{\pgfqpoint{0pt}{0pt}}
  \pgfpathlineto{\pgfqpoint{1.2pt}{1.2pt}}
  \pgfusepath{draw}
}

\newtcolorbox{kkbox}[1]{left=0.25mm, right=0.25mm, top=0.25mm, bottom=0.25mm, colframe=blue!66!black, boxrule=0.5pt, title={#1}, fonttitle=\bfseries, coltitle=blue!66!black, attach title to upper={\ }}

\usepackage{mdframed}
\mdfdefinestyle{MyFrame}{
	linecolor=black,
	backgroundcolor=gray!10,
	innertopmargin=5pt,
	innerbottommargin=5pt,
	innerrightmargin=5pt,
	innerleftmargin=5pt,
}

\usepackage[strict]{changepage}
\usepackage{framed}
\definecolor{customshade}{rgb}{0.941, 0.937, 0.996}
\definecolor{darkblue}{rgb}{0.14,0.22,0.62}
\newenvironment{kkboxline}{%
  \MakeFramed{\advance\hsize-\width\FrameRestore}%
  \noindent\hspace{-4.55pt}%
  \begin{adjustwidth}{}{7pt}%
}
{%
  \end{adjustwidth}\endMakeFramed%
}

\usepackage{listings}
\usepackage{microtype}
\usepackage{graphicx}
\usepackage{caption}
\usepackage{subcaption}
\usepackage{blindtext}
\usepackage{threeparttable}
\usepackage{enumitem}

\usepackage{multirow}
\usepackage{multicol}
\usepackage{nicefrac}
\usepackage{microtype}
\usepackage{lipsum}
\usepackage{graphicx}
\usepackage{float}
\usepackage{algorithm}
\usepackage[noend]{algpseudocode} %
\usepackage{booktabs}
\usepackage{makecell}
\usepackage{colortbl}
\usepackage{contour}
\usepackage{xspace}

\definecolor{Gray}{gray}{0.92}
\newcolumntype{g}{>{\columncolor{Gray}}c}
\newcolumntype{G}{>{\columncolor{Gray}}r}

\newcommand*\circled[1]{\tikz[baseline=(char.base)]{
            \node[shape=circle,draw,inner sep=0.3pt] (char) {#1};}}

\newcommand{\Tech}{{\textsc{SOFT}\xspace}}

\begin{document}

\date{}

\title{\Large \bf SOFT: Selective Data Obfuscation for Protecting LLM Fine-tuning \\ against Membership Inference Attacks}

\author{
  {\rm Kaiyuan Zhang, Siyuan Cheng, Hanxi Guo, Yuetian Chen, Zian Su, Shengwei An, Yuntao Du,}\\ {\rm Charles Fleming$^{\dagger}$, Ashish Kundu$^{\dagger}$, Xiangyu Zhang, Ninghui Li
  }\\
  {\rm Purdue University, $^{\dagger}$Cisco Research}\\
  \tt\normalsize \{zhan4057,\,cheng535,\,guo778,\,chen5264,\,su284,\,an93,\,ytdu,\,xyzhang,\,ninghui\}@purdue.edu,\\
  \tt\normalsize $^{\dagger}$\{chflemin,\,ashkundu\}@cisco.com \\
} %

\maketitle

\begin{abstract}
Large language models (LLMs) have achieved remarkable success and are widely adopted for diverse applications. However, fine-tuning these models often involves private or sensitive information, raising critical privacy concerns. 
In this work, we conduct the first comprehensive study evaluating the vulnerability of fine-tuned LLMs to membership inference attacks (MIAs).
Our empirical analysis demonstrates that MIAs exploit the loss reduction during fine-tuning, making them highly effective in revealing membership information.
These findings motivate the development of our defense.
We propose \Tech{} (\textbf{S}elective data \textbf{O}bfuscation in LLM
\textbf{F}ine-\textbf{T}uning), 
a novel defense technique that mitigates privacy leakage by leveraging influential data selection with an adjustable parameter to balance utility preservation and privacy protection. 
Our extensive experiments span six diverse domains and multiple LLM architectures and scales.
Results show that \Tech{} effectively reduces privacy risks while maintaining competitive model performance, offering a practical and scalable solution to safeguard sensitive information in fine-tuned LLMs.\footnote{Code is available at \href{https://github.com/KaiyuanZh/SOFT}{https://github.com/KaiyuanZh/SOFT}.}

\end{abstract}

\section{Introduction}\label{sec:intro}

Large language models are gaining significant public interest with diverse applications~\cite{gpt3, gpt4, claude, swebench, agents_anthropic}.
With LLMs continuously improving and model sizes growing, there is significant interest in understanding the potential privacy threat to the data used to train LLMs.  One common way to assess the privacy threat is to conduct \textit{membership inference attacks} (MIA)~\cite{shokri2017membership}, which determine whether a specific data record was used to train a target model or not. 

Researchers have empirically demonstrated that LLMs 
can be vulnerable to membership inference attacks~\cite{loss_mia, zlib_lowercase_ratio, min_k_prob, min_kpp, sok_mia_nowhere, recall_mia, con_recall}, reporting Area Under the Curve (AUC) values for MIA as high as 0.98~\cite{con_recall, recall_mia} in pre-trained LLMs. However, recent analyses have identified significant limitations in these findings.
Duan et al.~\cite{duan2024membership} and Maini et al.~\cite{maini2024llm} have attributed the high AUC of these membership inference attacks to the temporal shift between members and non-members in the Wiki-MIA dataset~\cite{min_k_prob}, rather than genuine membership leakage.
Recent works~\cite{zhang2024membership, blind_mia, sok_mia_nowhere} have further validated that existing MIAs for pre-training data are largely ineffective. 
This is not surprising, given that each piece of data is only used once~\cite{touvron2023llama,muennighoff2023scaling,komatsuzaki2019one} during LLMs pre-training.

While it is unclear that MIAs pose significant threats for the pre-training phase of LLMs, 
in real-world scenarios, LLMs are often \textit{fine-tuned} to be deployed across diverse downstream tasks.
As pre-training large-scale LLMs requires resources that are usually only available to large companies~\cite{gpt4, touvron2023llama, claude}, more and more small companies and individuals use pre-trained model as the backbone to fine-tune on downstream applications, such as medical~\cite{labrak2024biomistral}, clinical~\cite{jagannatha2021membership}, legal~\cite{colombo2024saullm}, code generation~\cite{su2024source,xu2024prosec,feng2025intentest}, and multilingual abilities~\cite{alves2024tower}, etc.
Data used in fine-tuning often includes either personally identifiable information (PII)~\cite{chen2024janus}, copyright data~\cite{liu2024rethinking}, or even confidential organizational information~\cite{nist_2024}, adding complexity to privacy and legal considerations.
The protection of such sensitive information sometimes falls under regulatory frameworks like the General Data Protection Regulation (GDPR)\cite{GDPR} in Europe and the California Consumer Privacy Act (CCPA)\cite{CCPA} in the United States, which establish guidelines for handling private information responsibly.

Recent studies have explored privacy leakage during the fine-tuning phase, including techniques such as prefix or prompt tuning~\cite{chen2024janus, spv_mia}, and adapters tuning~\cite{mireshghallah2022empirical_adapter}.
While these works have advanced our understanding of privacy risks in fine-tuned large language models, their scope remains limited. 
For example, some focus exclusively on smaller models, such as RoBERTa~\cite{liu2019roberta}, or employ compute-intensive fine-tuning techniques~\cite{spv_mia}.
Additionally, methods like Low-Rank Adaptation (LoRA)~\cite{hu2022lora} and comparative analyses of privacy leakage between full fine-tuning and alternative fine-tuning approaches in large language models remain underexplored.

This paper investigates privacy leakage in fine-tuned large language models (LLMs).  In addition to evaluating a wide range of existing MIA techniques, we also introduce an ensemble attack that combines features from existing attacks to better assess the extent of privacy risks. 
Using this ensemble attack and nine baseline MIAs, we conduct a comprehensive evaluation of privacy leakage in the Pythia suite of models~\cite{pythia}, trained on seven datasets from the Pile~\cite{pile_data}. 
Evaluation results from diverse datasets such as ArXiv, Wikipedia, and GitHub, reveal significant privacy risks in fine-tuned LLMs. 
For instance, most attacks achieve an AUC exceeding 0.8 when inferring membership status for Pythia-6.9B fine-tuned models.
To understand the mechanisms behind these privacy risks, we analyze the effectiveness of MIAs across datasets and model configurations. 
Existing MIAs can be categorized as reference-based or reference-free, and they mainly differ in strategies for distinguishing uncommon sentences used in training from common sentences excluded from training. 
Our empirical results demonstrate that reference-based attacks generally outperform reference-free ones.
We also find that privacy leakage increases with model size and data exposure, with significant risks evident even after one epoch of fine-tuning. 
LoRA offers better privacy protection than full fine-tuning but results in significant utility loss.

To mitigate privacy leakage from MIAs, existing defenses often modify training processes or model outputs. 
While methods like DP-LoRA~\cite{dpLoRA, private_transformer} offer differential privacy for LLM fine-tuning, they introduce memory overhead without achieving a favorable privacy-utility trade-off. 
This underscores the need for practical defenses tailored to fine-tuned LLMs that balance privacy protection and utility.
Building on insights gained from our analysis of MIAs, we propose \Tech{} (\textbf{S}elective data \textbf{O}bfuscation in LLM
\textbf{F}ine-\textbf{T}uning), a novel technique that mitigates privacy leakage by replacing influential samples in the fine-tuning dataset with obfuscated paraphrases. By targeting samples most vulnerable to MIAs, \Tech{} effectively balances performance and privacy.
In Section~\ref{sec:method}, we detail \Tech{}'s three-phase pipeline: warm-up fine-tuning, influential data selection, and data obfuscation. This iterative approach addresses privacy risks while maintaining scalability and practicality in LLM fine-tuning.

We summarize our contributions as follows:
\begin{itemize}
    \item We present the first systematic study evaluating the vulnerability of fine-tuned LLMs to MIA. While prior work focuses on pre-trained LLMs, our study fills the gap by analyzing privacy risks in fine-tuned models.
    \item We propose \Tech{}, a novel defense mechanism that mitigates membership leakage by replacing influential samples with obfuscated paraphrases. By refining the selection of influential data through loss-based prioritization, \Tech{} effectively balances privacy protection and model utility.
    \item We perform extensive experiments to analyze the factors influencing MIAs in fine-tuned LLMs across different dataset categories and various LLM architectures and scales. Our results provide insights into the relationship between dataset properties, model configurations, and privacy risks.
\end{itemize}

\noindent
\textbf{Organization.}
The rest of this paper is organized as follows. 
Section~\ref{sec:preliminaries} formulates the problem and provides background on membership inference attacks and defenses.
Section~\ref{sec:observations} introduces the ensemble attack, discusses five key findings, and analyzes insights.
Section~\ref{sec:method} presents the design and analysis of our proposed defense, \Tech{}.
Section~\ref{sec:evaluation} provides a comprehensive evaluation of \Tech{} against various MIAs, comparing it with state-of-the-art defenses.
Section~\ref{sec:related_work} reviews related work.
Section~\ref{sec:conclusion} concludes the paper.

\section{Preliminaries}\label{sec:preliminaries}

In this section, we start by giving our problem definition and threat model. 
Then we provide a brief overview of LLM fine-tuning, e.g. full fine-tuning and LoRA~\cite{hu2022lora, loravsFullfinetune}.
Finally, we introduce various existing MIA methods and mitigation techniques.

\subsection{Problem Definition}

In this paper, we focus on membership inference attack~\cite{shokri2017membership, carlini2022first_princeple, document_level_mia} within the context of fine-tuned LLMs. Assuming a dataset $\mathcal{D}$ is used to fine-tune a large language model $f_{\text{FT}}$.
Note that $\mathcal{D}$ is not necessarily disjoint from the pre-training dataset.
The objective of an MIA is to determine whether a given instance $x$ is part of $\mathcal{D}$ (i.e. $x \in \mathcal{D}$). 

\noindent
\textbf{Threat Model.}
Our threat model is consistent with the existing MIAs for LLMs in the literature~\cite{zlib_lowercase_ratio, min_k_prob, min_kpp, sok_mia_nowhere, recall_mia, con_recall}, while we focus on the \textbf{fine-tuning} phase of LLM.
Note that MIAs against the pre-training phase of LLMs have been proposed with varying assumptions made for the adversary, from access to the model weights~\cite{liu2024probing}, to only
query access with model logits~\cite{zlib_lowercase_ratio,min_k_prob, min_kpp, recall_mia, document_level_mia, con_recall, sok_mia_nowhere} or merely generated text~\cite{duarte2024cop}.
Some work also assumes access to additional nonmember data from the same distribution as the target sample~\cite{recall_mia, con_recall, liu2024probing}.
We assume the adversary can query both the pre-trained model and the target fine-tuned model to get model logits and data distribution.
This scenario aligns with common deployment settings for commercial LLMs, where the adversary merely obtains the label or the predicted words along with their associated probabilities.

\subsection{Fine-tuning LLMs}~\label{subsec:prelim_fine-tune}
Directly using large models for specific tasks often results in suboptimal performance. As a result, fine-tuning has become an essential approach when adapting pre-trained language models to downstream tasks.
Full fine-tuning, which updates all the model parameters, has been a longstanding practice in the NLP domain.
However, with the advent of scaling laws~\cite{scaling_laws} and increasingly large foundation models~\cite{gpt3, gpt4, meta2024llama}, Parameter-efficient fine-Tuning (PEFT) has gained more attention.
Among PEFT methods, LoRA has become the \textit{de-facto} PEFT approach~\cite{xu2023parameter} due to its simplicity and effectiveness, and is widely adopted in real-world popular LLMs~\cite{zhao2024lora} and libraries, such as Hugging Face PEFT and Databricks~\cite{lora_databricks}.
In this work, we focus on examining privacy leakage risks during the LLMs fine-tuning, with emphasis on full fine-tuning and LoRA.

\begin{figure}[ht]
    \centering
    \begin{minipage}[t]{0.235\textwidth}
        \centering
        \includegraphics[width=0.95\textwidth]{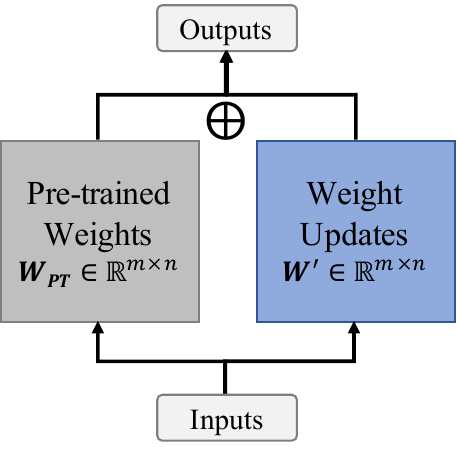}
        \subcaption{Full Fine-tuning}
    \end{minipage}
    \hfill
    \begin{minipage}[t]{0.235\textwidth}
        \centering
        \includegraphics[width=0.95\textwidth]{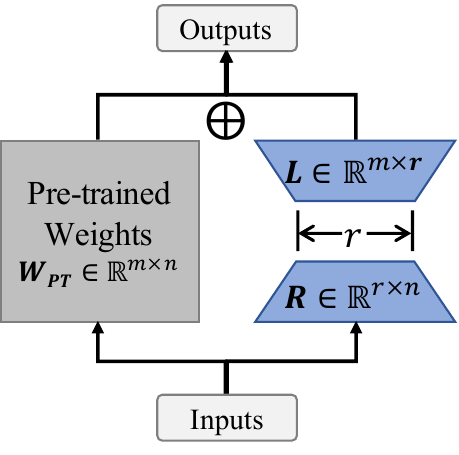}
        \subcaption{LoRA}
    \end{minipage}
    \caption{Full Fine-tuning vs. LoRA.}
    \label{fig:full_vs_lora}
\end{figure}

\noindent
\textbf{Fine-tuning.}
Given a pre-trained language model $f(W_{\text{PT}}; x)$ parameterized by $W_{\text{PT}} \in \mathbb{R}^{m\times n}$, where $x$ is any input,
the fine-tuned model is represented as:
\begin{equation}\label{equation:fine_tune}
    f_{\text{FT}}(W_{\text{PT}} \oplus \Delta W; x),
\end{equation}
where %
$\Delta W \in \mathbb{R}^{m\times n}$ represents the additional parameters learned during fine-tuning.

In the following, we instantiate fine-tuning in both \textit{full fine-tuning} and \textit{fine-tuning via LoRA}.

\noindent
\textbf{Full Fine-tuning.} 
In \textit{full fine-tuning}, all model parameters are updated, such that $\Delta W = W'$, 
where $W' \in \mathbb{R}^{m\times n}$. 
This approach updates \textit{all} parameters of the pre-trained weights, modifying $W_{\text{PT}}$ in its full dimensionality to obtain the fine-tuned model.

\noindent
\textbf{Fine-tuning via LoRA.}
While full fine-tuning treats all parameters as trainable, LoRA~\cite{hu2022lora, loravsFullfinetune} reduces the parameter number by reparameterizing the weight updates $\Delta W$ as the product of two low-rank matrices, such that $\Delta W = LR$, 
where  $L\in \mathbb{R}^{m\times r}$, $R \in \mathbb{R}^{r\times n}$, and the rank $r \ll min(m, n)$.
where $W_{\text{PT}}$ remains frozen, and the learned updates $LR$ are added to the pre-trained weights.

\noindent
\textbf{Comparison of Full Fine-Tuning and LoRA.}
The number of trainable parameters per weight matrix is $mn$ in full fine-tuning, and $mr + rn$ in LoRA.  
When the rank $r$ is small, LoRA's reparameterization significantly reduces the computational cost.

\subsection{Membership Inference Attacks in LLMs}\label{subsec:mia_existing_attacks}

Prior MIAs mainly focused on classical machine learning
models, such as classification models~\cite{shokri2017membership,carlini2022first_princeple, choquette2021label, liu2022membership}.
Recently, more and more methods~\cite{min_k_prob,min_kpp,recall_mia,con_recall} have been proposed to perform MIAs against pre-trained LLMs.

Given a model $\mathcal{M}$, an MIA is given by a membership scoring function $s(x; \mathcal{M})$, such that for each instance $x$, a higher $s(x; \mathcal{M})$ value means that the MIA considers that $x$ is more likely to be a member. 

\noindent
\textbf{Loss.} Many existing MIAs against classification models use the loss of an instance to determine membership~\cite{loss_mia}.  Since LLM training also aims to reduce the loss of the training instances, it is natural to use 
the model's computed negative loss over the target sample as the membership score. 
Denote the score function as $\ell$, then we have:
\begin{equation}
    s(x; \mathcal{M}) = -\ell(x, \mathcal{M}).
\end{equation}
We define the loss function based on the negative log-likelihood as,
\begin{equation}
    \ell(\theta; x) = - \sum_{i=1}^n \log p(x_i \mid x_1, \ldots, x_{i-1}; \theta),
\end{equation}
where $p$ denotes the conditional probability of $x_i$ given the preceding inputs, parameterized by $\theta$.

The main drawback of directly using loss is that a non-member common sentence may have a lower loss than an uncommon sentence that is used in training.  Many other MIAs have been proposed to address this by using different calibration approaches.

\begin{kkboxline}
\textbf{The Calibration Challenge.}
\textit{Existing LLM MIAs mainly differ on how to differentiate \textbf{uncommon} sentences used in training from \textbf{common} sentences not used in training.
Many of these methods share similarities on \textbf{calibration} and differ mainly in their use of loss, log-likelihood, perplexity, contrastive ratios, or an extra reference model.
}
\end{kkboxline}

\noindent

\noindent
\textbf{Zlib.}
 Zlib Entropy~\cite{zlib_lowercase_ratio} computes the ratio of the target sample loss and the bit length of $x$ after zlib compression~\cite{gailly2004zlib}, i.e. zlib$(x)$.
 Essentially, this approach uses zlib compression length of a sentence as an approximation of whether the sentence is common, and uses that to normalize loss for the purpose of membership inference. 
 The score function is defined as:
 \begin{equation}
     s(x; \mathcal{M}) = \frac{-\ell(x, \mathcal{M})}{\text{zlib}(x)}.
 \end{equation}

\noindent
\textbf{Lowercase.} 
Lowercase~\cite{zlib_lowercase_ratio} normalizes the loss of the target sample $x$ by dividing it by the loss of its \textit{lowercased} version $x_{lowercase}$.  
The score function is defined as:
\begin{equation}
    s(x; \mathcal{M}) = \frac{\ell(x, \mathcal{M})}{\ell(x_{lowercase}, \mathcal{M})}.
\end{equation}
The intuition is that if $x$ was used in training, then the model has optimized for the specific casing in $x$, and this ratio is likely to be higher than when $x$ was not used in training.

\noindent
\textbf{Min-K\% Prob.}
Min-K\% probability~\cite{min_k_prob} deals with the calibration challenge by focusing on the $k$ tokens that have the lowest predicted probabilities.  
More specifically, the membership score is defined as the average log-likelihood of these minimum-probability tokens, as follows:
 \begin{equation}
     s(x; \mathcal{M}) = \frac{1}{\abs{\min\text{-}k(x)}} \sum_{x_i \in \abs{\min\text{-}k(x)}} - \log p(x_i \mid x_1, \cdots, x_{i-1}).
 \end{equation}
The intuition is that the score will not be affected by having many common tokens in a sentence. If an instance has been used in the training, even the $k$ least common tokens in the instance may have higher predicted probabilities.  This approach, however, still cannot correctly identify non-members that have very few uncommon tokens.

\noindent
\textbf{Min-K\%++.} 
Min-K\%++~\cite{min_kpp} extends Min-K\% by calibrating the log-likelihood of each token sequence using the mean and standard deviation over the vocabulary of the model, defined as:
\begin{equation}
    s_{\text{token}}(x_{<t}, x_t; \mathcal{M}) = \frac{\log p(x_t \mid x_{<t}; \mathcal{M}) - \mu_{x<t}}{\sigma_{x<t}},
\end{equation}
where \(x_{<t}\) is the prefix. \(\mu_{x<t}\) and \(\sigma_{x<t}\) denote the mean and standard deviation of the log-likelihoods over the model's vocabulary, respectively.
\begin{equation}
    s(x; \mathcal{M}) = \frac{1}{|\min\text{-}k(x)|} \sum_{(x_{<t}, x_t) \in \min\text{-}k(x)} s_{\text{token}}(x_{<t}, x_t; \mathcal{M}),
\end{equation}
where \(\min\text{-}k(x)\) denotes the set of token sequences corresponding to the \(k\%\) lowest token-level scores for the input sequence \(x\).

\noindent
\textbf{Ratio.} 
Ratio~\cite{zlib_lowercase_ratio} determines membership by comparing the loss ratio between a target model and a reference model (or shadow model) on a target sample.
The reference model serves as a calibration baseline.
The intuition behind Ratio is that members are expected to have a lower loss on the target model $\mathcal{M}$ than the reference model $\mathcal{R}$.
The membership score is defined as:
\begin{equation}
    s(x; \mathcal{M}) = \frac{\ell(x; \mathcal{M})}{\ell(x; \mathcal{R})}.
\end{equation}

\noindent
\textbf{ReCall.} 
ReCall~\cite{recall_mia} quantifies the log-likelihood of a target sample \(x\) with a non-member prefix \(P_{\text{non-member}}\) context.
The method assumes access to a small number of non-member samples. 
The intuition is that introducing a non-member context \(P_{\text{non-member}}\) alters the likelihood of a sample depending on its membership status. For non-members, the prefix introduces a new, unseen context that significantly affects the likelihood, while for members, the impact is smaller.
The score is defined as:
\begin{equation}
    s(x; \mathcal{M}) = \frac{\log p(x \mid P_{\text{non-member}}; \mathcal{M})}{\log p(x; \mathcal{M})}.
\end{equation}

\noindent
\textbf{CON-ReCall.} 
CON-ReCall~\cite{con_recall} leverages contrastive decoding to enhance membership inference by comparing the log-likelihoods of a target text \(x\) under both member and non-member prefixes. 
The method assumes access to both member and non-member samples.
This approach builds on the observation that member and non-member prefixes induce asymmetric shifts in likelihood, using these differences to determine membership.
The membership score is defined as:
\begin{equation}
    s(x; \mathcal{M}) = \frac{\log p(x \mid P_{\text{non-member}}; \mathcal{M}) - \gamma \cdot \log p(x \mid P_{\text{member}}; \mathcal{M})}{\log p(x; \mathcal{M})},
\end{equation}
where \(\gamma\) controls the contrast strength. \(P_{\text{member}}\) and \(P_{\text{non-member}}\) are prefixes composed of member and non-member contexts respectively.

\noindent
\textbf{Bag of Words (BoW).}
Recent studies~\cite{duan2024membership,blind_mia,sok_mia_nowhere} have demonstrated that MIA is largely ineffective against pre-trained LLMs.
A Bag of Words~\cite{sok_mia_nowhere} classifier \(f\) is solely trained on dataset features to distinguish members from non-members, without considering the model:
\begin{equation}
    s(x;\mathcal{M}) = f(v(x)),
\end{equation}
where \(v(x)\) denotes corresponding textual features. 
An AUC significantly exceeding 0.5 when evaluated with BoW indicates a pronounced distribution shift in the dataset rather than actual data leakage. Similar findings are also reported by~\cite{blind_mia}.

\subsection{Membership Inference Defenses in LLMs}
Existing defenses are mostly grounded on differential privacy (DP), focusing on either pre-training~\cite{anil2021large,hoory2021learning} or fine-tuning~\cite{dpLoRA,private_transformer} with privacy guarantees. 
Next, we introduce several current defense mechanisms against MIAs in the LLM context.

\noindent
\textbf{DP.}
DP~\cite{dp2006, dp2014} is a formal privacy definition for algorithms that operate on sensitive datasets.
An algorithm is differentially private if and only if 
including or excluding any individual sample does not substantially influence the distribution of algorithm outputs.

\begin{definition} [$(\epsilon, \delta)$-Differential Privacy]\label{def:dp_defi}
A randomized algorithm $\mathcal{F}: \mathcal{X} \rightarrow \mathcal{Y}$ is $(\epsilon, \delta)$-differentially private if for all adjacent datasets $X$, $X' \in \mathcal{X}$ and all $Y \subseteq \mathcal{Y}$, it holds that
\[
\Pr[\mathcal{F}(X) \in Y] \leq \text{exp}(\epsilon) \Pr[\mathcal{F}(X') \in Y] + \delta,
\]
where $\epsilon$ and  $\delta$ denote the privacy budget and the privacy loss parameter, respectively.
\end{definition}

\noindent
\textbf{Fine-tuning via DP-SGD.} Differentially private stochastic gradient descent (DP-SGD)~\cite{song2013stochastic, bassily2014private, dpSGD} is the most widely adopted algorithm for training machine learning models with privacy guarantees. 
In this paper, we focus on the application of DP-SGD in the context of fine-tuning.

\noindent
\textbf{Fine-tuning via DP-LoRA.} DP-LoRA~\cite{dpLoRA} is a notable adaptation of DP-SGD specifically designed for fine-tuning large language models (LLMs).
Recall LoRA in Section~\ref{subsec:prelim_fine-tune},
fine-tuning with DP-LoRA restricts modified gradient updates to the adaptation layers (LoRA weights).
Compared with DP-SGD in LLMs, DP-LoRA reduces the memory and increases the training speed~\cite{dpLoRA},
while still unable to scale large datasets.

\section{Privacy Leak Analysis in Fine-tuned LLMs}\label{sec:observations}

In this section, we first propose a simple yet effective ensemble attack that leverages existing membership inference attack features to enhance the analysis of privacy leakage. Using the ensemble attack and other MIA baselines, we then investigate privacy leakage from five distinct perspectives. The insights derived from this analysis inform the design of our proposed defense, \Tech{}, which is introduced in the subsequent section.

\subsection{Ensemble Attack}
Building upon the existing MIAs discussed in Section~\ref{subsec:mia_existing_attacks} and inspired from~\cite{maini2024llm}, we introduce an ensemble attack that combines multiple MIAs to enhance membership inference accuracy.

As outlined in ``\textit{The Calibration Challenge},'' existing LLM MIAs primarily differ in their approach to distinguishing uncommon sentences used in training from common sentences not included in training. These attacks can be broadly categorized into two groups: (1) reference-based attacks and (2) reference-free attacks. For instance, Ratio~\cite{zlib_lowercase_ratio} is a reference-based attack that determines membership by comparing the loss ratio between a target model and a reference model. 
In contrast, all other attacks are reference-free, including Loss~\cite{loss_mia}, Zlib~\cite{zlib_lowercase_ratio}, Lowercase~\cite{zlib_lowercase_ratio}, Min-K\% Prob~\cite{min_k_prob}, Min-K\%++~\cite{min_kpp}, Bag of Words~\cite{sok_mia_nowhere}, ReCall~\cite{recall_mia}, and CON-ReCall~\cite{con_recall}. We evaluate these nine MIAs using the Pythia~\cite{pythia} suite of models (default as Pythia-6.9B), e.g. 70M, 160M, 410M, 1B, 1.4B, 2.8B, 6.9B, on seven various subsets of the Pile dataset~\cite{pile_data}, such as ArXiv, DeepMind Mathematics, HackerNews, PubMed, Pile CC, Wikipedia, and GitHub. The dataset partitioning follows~\cite{duan2024membership}, and the evaluation is conducted with 13-grams data split (13\_0.8) by default.

\begin{figure}[ht]
    \centering
    \includegraphics[width=1\linewidth]{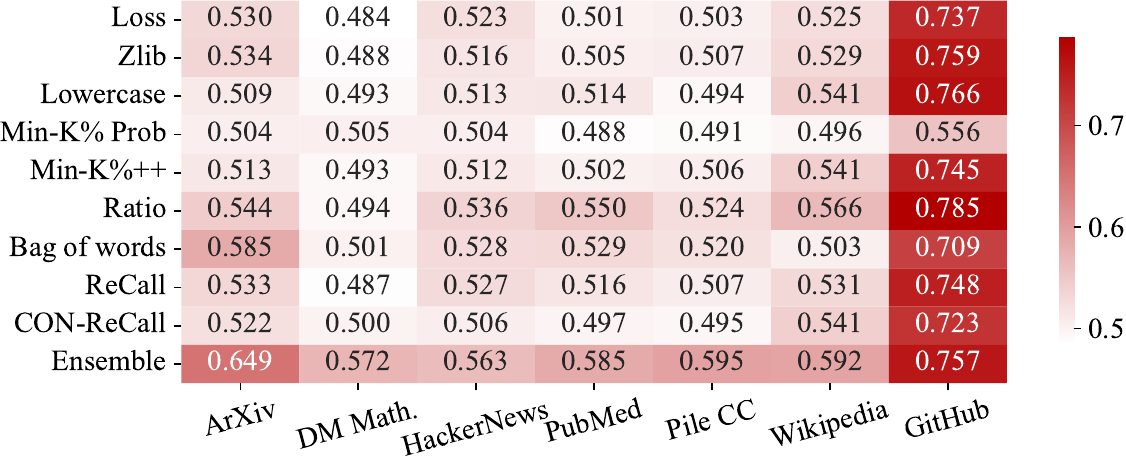}
    \caption{AUC-ROC of MIAs on Pre-trained Pythia-6.9B.}
    \label{fig:heatmap_13_gram_pretrain_pythia}
\end{figure}

\noindent
\textbf{Ensemble Attack.}
An effective MIA should achieve an AUC significantly greater than 0.5. In Figure~\ref{fig:heatmap_13_gram_pretrain_pythia}, we observe that no single MIA consistently outperforms others across all datasets, except for the ensemble approach in the last row. 
However, if there exists MIAs demonstrate performance slightly better than random guessing, we can selectively combine those providing positive signals and aggregate their features to train a classifier tailored to a specific dataset. 
For the \textit{Ensemble} attack, we strategically select twelve features: loss, perplexity, lowercase, zlib, seven features derived from Min-K\%++ with varying $k$ thresholds, and ratio. The last row in Figure~\ref{fig:heatmap_13_gram_pretrain_pythia} demonstrates that the Ensemble attack significantly outperforms other MIAs across diverse scenarios.

\noindent
\textbf{Remark.}
The ineffectiveness of existing membership inference attacks in pre-trained LLMs, motivating the proposal of the \textit{Ensemble} attack.

\subsection{Pitfalls in Full Fine-tuning}
While several studies~\cite{lora_databricks, loravsFullfinetune, zhao2024lora, hu2022lora} provide valuable insights into the performance of full fine-tuning and LoRA across various domains and datasets, research on privacy leakage in fine-tuned LLMs remains limited. Although some findings may seem intuitive at first glance, they often stem from the intricate interplay of attack designs (e.g., loss-based attacks) and data properties (e.g., code, math).

\begin{kkboxline}
\textbf{Finding 1:}
\textit{
As model size and fine-tune epoch increase, fully fine-tuned LLMs exhibit greater privacy leakage. Even one-epoch fine-tuning results in significant leakage.
}
\end{kkboxline}

\noindent
\textbf{Description.}
We evaluate the Pythia~\cite{pythia} family of models trained on the Pile~\cite{pile_data} dataset, covering six model sizes: 70M, 160M, 1B, 1.4B, 2.8B, and 6.9B parameters, and fine-tune them on seven distinct downstream datasets. To investigate the effect of data exposure, we fine-tune Pythia-6.9B for varying numbers of epochs across the same downstream datasets. 
To ensure a fair comparison, we maintain consistent core training settings, including the optimizer, learning rate schedule, total training steps, and other hyperparameters.

\begin{figure}[ht]
    \centering
    \includegraphics[width=1\linewidth]{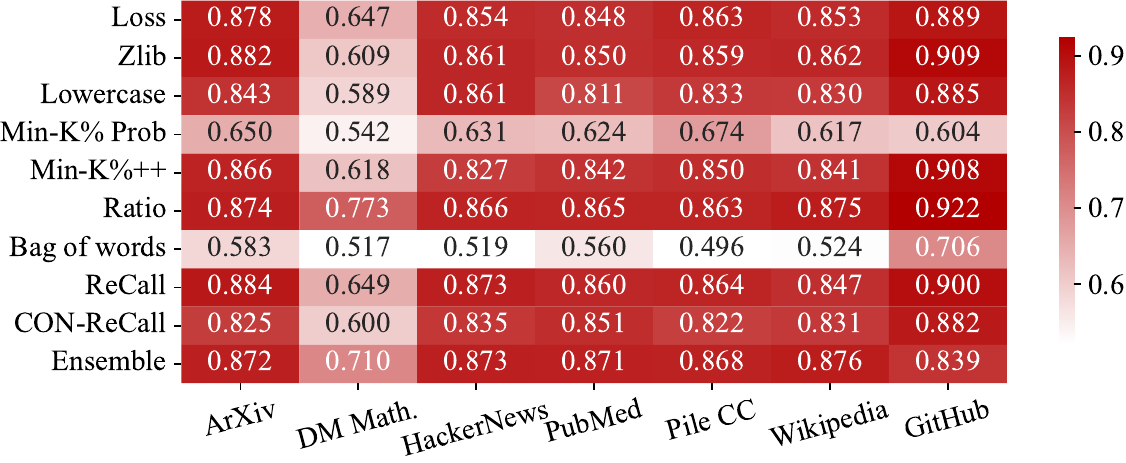}
    \caption{AUC-ROC on Full Fine-tuned Pythia (3 Epochs).}
    \label{fig:heatmap_13_gram_full_finetune_pythia_epoch3}
\end{figure}

\begin{figure}[ht]
    \centering
    \begin{minipage}[t]{0.47\textwidth}
        \centering
        \includegraphics[width=1.0\textwidth]{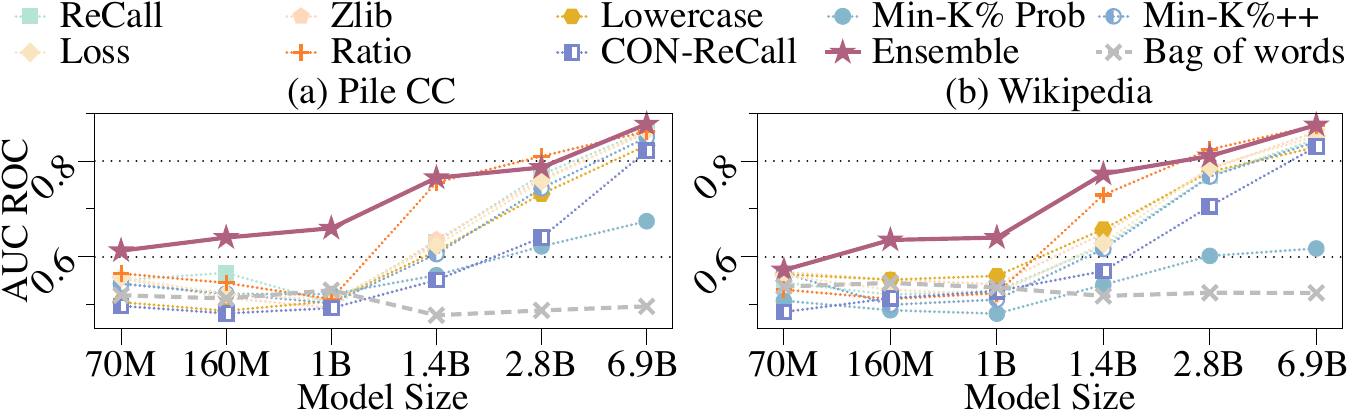}
        \subcaption{Full Fine-tune on Different Model Sizes of Pythia (3 Epochs).}
        \label{fig:model-size}
    \end{minipage}

    \begin{minipage}[t]{0.47\textwidth}
        \centering
        \includegraphics[width=1.0\textwidth]{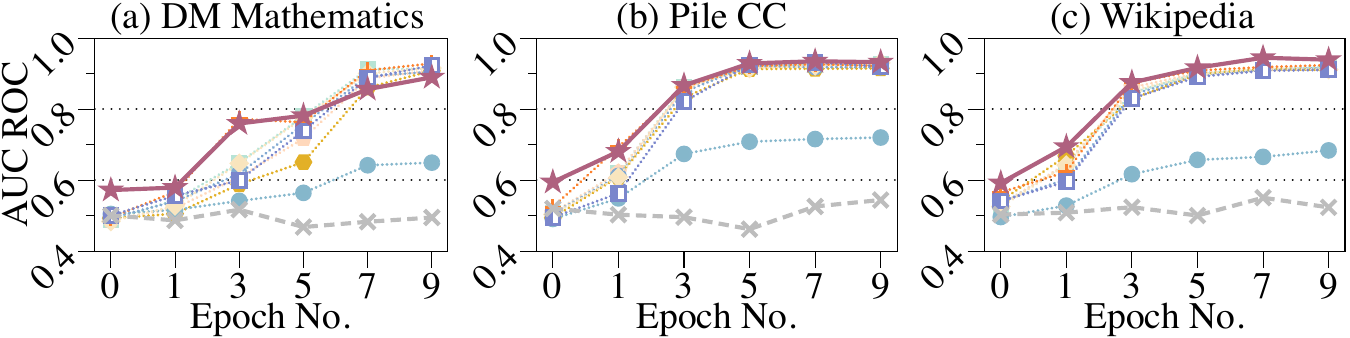}
            \subcaption{Full Fine-tune with Different Epochs on Pythia-6.9B.}
        \label{fig:epoch}
    \end{minipage}
    \caption{AUC-ROC of MIAs against Full Fine-tuned Models.}
\end{figure}

\noindent
\textbf{Insights.}
Our observations indicate that larger models consistently enhance membership inference attack (MIA) performance, with a clear upward trend in full fine-tuning scenarios, as shown in Figure~\ref{fig:model-size}. Since it is a standard practice to pre-train LLMs for approximately one epoch due to the scale of the training data~\cite{touvron2023llama, muennighoff2023scaling, komatsuzaki2019one}. 
For most existing MIAs, the performance on pre-trained LLMs is close to \textit{random guessing}. However, after full fine-tuning, some of these attacks can achieve AUC scores as high as 0.9 regardless of domains, as illustrated in Figure~\ref{fig:heatmap_13_gram_full_finetune_pythia_epoch3}. 

Regarding fine-tune epochs, we observe a significant boost in attack performance even with only one epoch of full fine-tuning. Figure~\ref{fig:epoch} demonstrates the increasing AUC performance for various MIAs as the number of fine-tuning epochs grows. 
Notably, the Bag of Words attack is useful for detecting distribution shifts in datasets, with an AUC around 0.5 indicating a fair dataset distribution.

\noindent
\textbf{Remark.}
In summary, full fine-tuning demonstrates significant privacy leakage, which motivates our defense proposal, \Tech{}.

\subsection{Privacy-Utility Trade-offs in LoRA}
\begin{kkboxline}
\textbf{Finding 2:} 
\textit{LoRA with different ranks provides privacy-utility trade-offs.  
Overall, LoRA demonstrates better privacy protection compared to full fine-tuning.}
\end{kkboxline}

\noindent
\textbf{Description.}
We fine-tune the Pythia-6.9B model using LoRA for 5 epochs, as LoRA generally requires more time to converge compared to full fine-tuning. The ranks evaluated include 1, 8, 32, 64, and 128. 
Empirically, setting $\alpha = 2r$ has been established as a common practice in LoRA~\cite{lora_databricks}, where $\alpha$ is the scaling factor and $r$ denotes the rank. We adopt this parameterization as the default configuration. 
Our evaluation ensures fair comparisons by converging both full fine-tuning and LoRA models to the same performance level.

\begin{figure}[ht]
    \centering
    \includegraphics[width=1\linewidth]{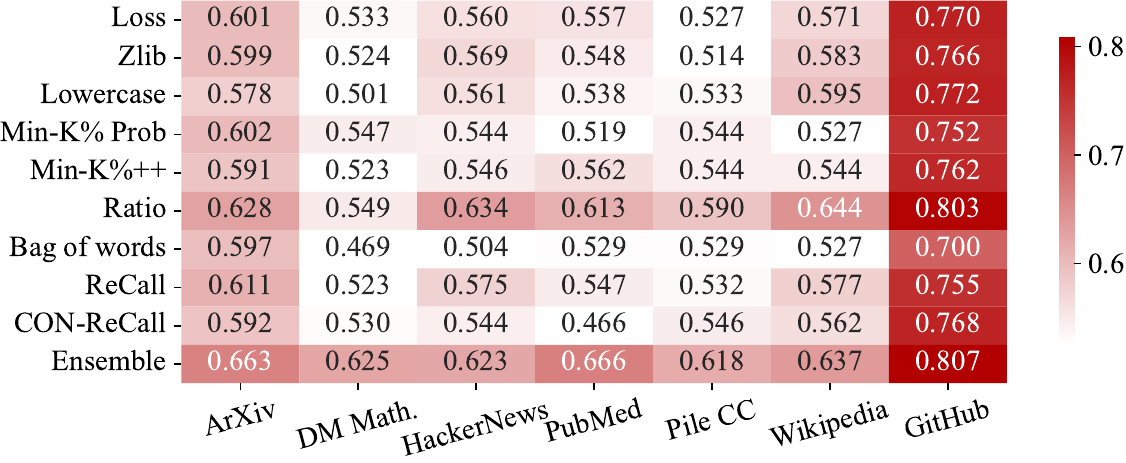}
    \caption{AUC-ROC on LoRA Fine-tuned Pythia (5 Epochs).}
    \label{fig:heatmap_13_gram_LoRA_pythia_epoch5}
\end{figure}
\begin{figure}[!ht]
    \centering
    \includegraphics[width=1\linewidth]{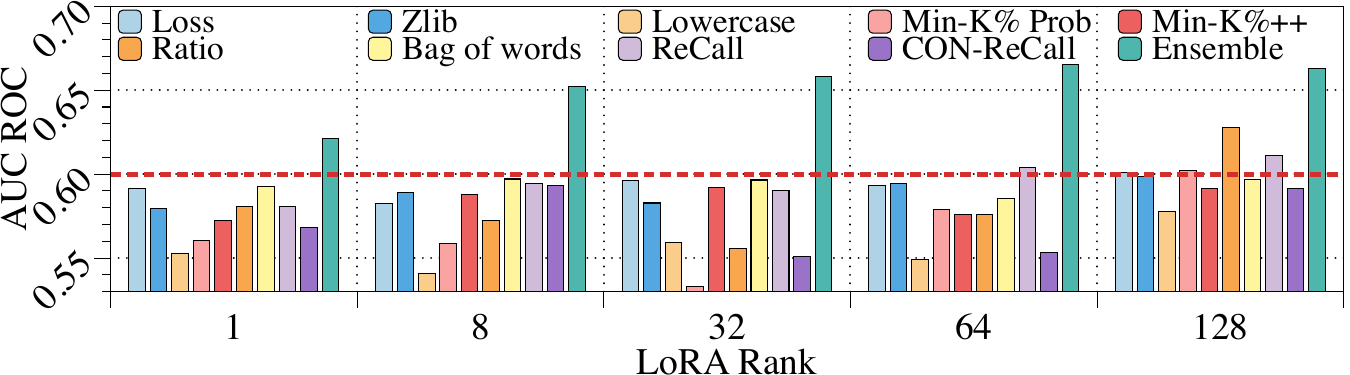}
    \caption{AUC-ROC on LoRA Fine-tuned in Different Ranks.}
    \label{fig:lora_rank_ablation}
\end{figure}

\noindent
\textbf{Insights.}
LoRA provides a favorable privacy-utility trade-off, fundamentally operating as a form of continual learning~\cite{continual_learning}, where a subset of parameters is fine-tuned to adapt the base model's capabilities to new domains. 
We observe that vanilla LoRA fine-tuning provides a balance between model performance and privacy protection, as evidenced by comparing Figure~\ref{fig:heatmap_13_gram_full_finetune_pythia_epoch3} (full fine-tuning) and Figure~\ref{fig:heatmap_13_gram_LoRA_pythia_epoch5} (LoRA). 
However, identifying this privacy-utility trade-off requires careful tuning. 
Recent works~\cite{lora_databricks, adapter_tuning} demonstrate that higher ranks and proper hyperparameter configurations significantly improve LoRA's performance. 
Our study corroborates these findings, revealing that higher ranks can exacerbate privacy leakage, as shown in Figure~\ref{fig:lora_rank_ablation}.

\noindent
\textbf{Interpretation of the LoRA Privacy-Utility Trade-off.} 
LoRA and full fine-tuning produce structurally different parameter updates, characterized by \textit{intruder dimensions}~\cite{loravsFullfinetune}, which are absent in full fine-tuning. 
Intuitively, intruder dimensions correspond to high-ranking singular vectors in weight matrices with large associated singular values.
When fine-tuning with LoRA, these intruder dimensions emerge and are approximately orthogonal to the singular vectors in the pre-trained weight matrix. 
In contrast, full fine-tuning maintains spectral similarity to the pre-trained model and does not introduce intruder dimensions. 
The existence of intruder dimensions in LoRA fine-tuning complicates its ability to achieve performance equivalent to full fine-tuning on a given task, explaining the observed trade-off in privacy and utility.

\noindent
\textbf{Remark.}
While LoRA achieves a trade-off between privacy and utility, the ensemble and ratio attack remain capable of compromising it, motivating the proposal of our defense, \Tech{}.

\subsection{Dataset Properties}
\begin{kkboxline}
\textbf{Finding 3:}
\textit{
Dataset properties significantly affect MIA difficulty. 
}
\end{kkboxline}

\noindent
\textbf{Description.}
State-of-the-art pre-trained LLMs are trained on billions to trillions of tokens~\cite{touvron2023llama, meta2024llama}, with dataset domains exhibiting varying properties. 
These variations affect the difficulty of performing MIAs.
We follow the experimental setup described in \textit{Findings 1 and 2}, fine-tuning the models using both full fine-tuning and LoRA for 3 and 5 epochs, respectively, to evaluate the impact of dataset properties.

\noindent
\textbf{Insights.}
While a few studies~\cite{lora_databricks, dubois2024alpacafarm} have rigorously compared full fine-tuning and LoRA across various datasets, including challenging domains such as code and mathematics, our study focuses on the privacy implications of dataset properties.
In Figure~\ref{fig:heatmap_13_gram_full_finetune_pythia_epoch3}, we observe that mathematical datasets exhibit lower AUC scores compared to other domains. 
Additionally, the GitHub dataset demonstrates significant distribution shifts, as evidenced by the Bag of Words~\cite{sok_mia_nowhere} AUC exceeding 0.5. This high overlap is attributable to the repetitive nature of code, including function definitions, code structures, and commonly used frameworks and libraries.

\noindent
\textbf{Remark.}
Given that different datasets and samples possess specialized properties, we strategically design an influence function to estimate the impact of individual samples in our proposed defense, \Tech{}.

\subsection{Attack Properties}
\begin{kkboxline}
\textbf{Finding 4:}
\textit{
Reference-based attacks generally achieve the best performance, although there are specific settings other attacks perform better.
}
\end{kkboxline}

\noindent
\textbf{Description.}
Using the experimental setup from \textit{Finding 1}, we conduct full fine-tuning with Pythia-6.9B, employing the 13-grams (13\_0.8) data partition by default across seven diverse datasets. 
In our evaluation, Ensemble and Ratio~\cite{zlib_lowercase_ratio} are reference-based attacks, while the remaining attacks are reference-free. The results highlight the consistent superiority of reference-based attacks, with certain scenarios favoring reference-free approaches.

\noindent
\textbf{Insights.}
Reference-based attacks rely on an auxiliary model trained on a dataset similar to the target model's training data for calibration. By comparing the outputs of the target and reference models, these attacks identify loss discrepancies that indicate whether a specific data point was part of the training set. 
As shown in Figure~\ref{fig:heatmap_13_gram_pretrain_pythia} (pre-trained LLMs) and Figure~\ref{fig:heatmap_13_gram_LoRA_pythia_epoch5} (LoRA fine-tuned LLMs), reference-based attacks, such as Ensemble and Ratio, consistently achieve the best performance. This approach leverages the observation that the reference model's loss on unseen data will differ from that of the target model if the data point was included in the training set.
Other than reference-based attacks, many MIA methods are loss-based or loss-variants, sharing similarities on calibration and differ mainly in their use of loss.

\noindent
\textbf{Remark.}
Building on these observations, our defense, \Tech{}, strategically leverages loss metrics to identify and prioritize influential data for enhanced privacy protection.

\subsection{Mitigation Strategies}
\begin{kkboxline}
\textbf{Finding 5:}
\textit{
DP-SGD and DP-LoRA add noise to each sample and cause a degradation in model performance.
}
\end{kkboxline}

\noindent
\textbf{Description.}
DP-SGD enforces differential privacy by bounding each sample's contribution to the gradient through per-sample gradient clipping and the addition of calibrated noise. This ensures privacy at the level of every individual training examples.
Similarly, DP-LoRA adapts LoRA to integrate differential privacy by applying clipping and noise addition to updates associated with individual samples.

\noindent
\textbf{Insights.}
In most scenarios, DP-LoRA~\cite{dpLoRA} demonstrates superior defensive performance. However, differential privacy techniques often result in notable utility degradation, particularly in large-scale LLM fine-tuning tasks, as empirically validated in Section~\ref{sec:evaluation}.
Besides, as noted in \textit{Finding 2}, LoRA fine-tuning can also achieve a reasonable balance between model performance and protection against MIAs.

\noindent
\textbf{Remark.}
DP-SGD and DP-LoRA add protection to all samples, however, it is feasible to identify individual vulnerable samples,  which motivates our \Tech{}.

\section{Proposed Defense: Selective Data Obfuscation by Paraphrasing} \label{sec:method}

\begin{figure*}[!ht]
    \centering
    \includegraphics[width=1\linewidth]{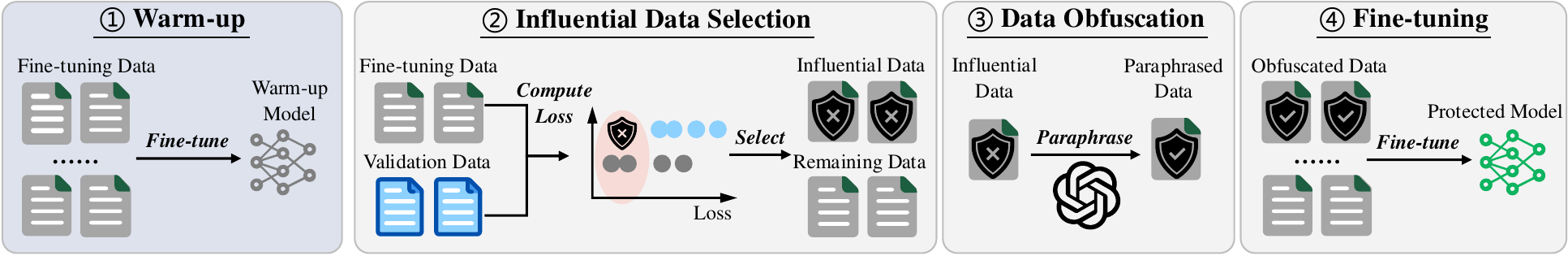}
    \caption{Overview of \Tech{}
    }
    \label{fig:overview}
\end{figure*}

Motivated from Section~\ref{sec:observations} findings, in this section, we propose a novel defense, \Tech{}.
We start by providing an overview of \Tech{}.
Then we introduce the details of \Tech{}, which paraphrases selected influential data.
Intuitively, fine-tuning on paraphrased data reduces the model's overfitting to the exact member samples, as it has not seen the raw member data, which can fool all the loss based MIAs.

\subsection{Overview of \Tech{}}

In high-level, \Tech{} involves substituting influential samples with semantically equivalent alternatives by a paraphraser during fine-tuning.
Inspired by \textit{influence functions}~\cite{ koh2017understanding},
we define influential samples as those vulnerable to MIA.
By doing so, \Tech{} provides a favorable balance between mitigating privacy leakage and maintaining fine-tuning performance.

\begin{definition}[Influence Function]
Influence functions quantify the effect of a single training point on model parameters or predictions. For a model $\mathcal{M}$ parameterized by $\theta$, the influence function measures the change in parameters $\theta$ when a training point $x_f$ is upweighted by an infinitesimal amount $\epsilon$:
\begin{equation}
I_{\text{params}}(x_f) = \frac{d\theta_{\epsilon}}{d\epsilon} \bigg|_{\epsilon=0} = -H^{-1}_{\theta} \nabla \ell(\mathcal{M}, x_f),
\end{equation}
where $H_{\theta} = \nabla^2 \mathcal{L}(\theta)$ is the Hessian of the total loss $\mathcal{L}(\theta)$, and $\nabla \ell(\mathcal{M}, x_f)$ is the gradient of the loss with respect to model parameters. 
\end{definition}
This formulation has been widely used to analyze the impact of training points on model behavior~\cite{koh2017understanding}.
While gradient-based influence is theoretically grounded~\cite{pruthi2020estimating}, \Tech{} employs a loss-based approximation for practical efficiency and direct alignment with the data selection mechanism.

Figure~\ref{fig:overview} outlines the pipeline of \Tech{}, which is divided into four main phases.
\underline{\circled{1} Warm-up} involves training the model on the entire dataset once, resulting in a warm-up model that help assess the sensitivity of each sample.
\underline{\circled{2} Influential Data Selection} entails computing the loss for both the validation and fine-tuning datasets during each subsequent fine-tuning iteration. Samples with losses below a specified threshold (typically set as the average validation loss) are selected as influential data.
\underline{\circled{3} Data Obfuscation} consists of paraphrasing the selected influential samples to obfuscate them and incorporating these obfuscated samples back into the fine-tuning process. \Tech{} updates the dataset by replacing the original samples with their obfuscated versions while retaining the remaining data.
\underline{\circled{4} Fine-tuning} fine-tunes the model using the updated dataset, thereby minimizing potential privacy leakage while preserving model utility.
Note that except for the warm-up phase, the remaining three phases are iterative and occur across multiple epochs.

\begin{algorithm}[!ht]
\small
\caption{Pseudocode of \Tech{}}
\label{algo:our_tech}
\begin{algorithmic}[1]
    \State \textbf{Input:} Pre-trained model $\mathcal{M}_{PT}$, fine-tuning dataset $\mathcal{D}_f$, validation dataset $\mathcal{D}_v$, paraphrasing strength $\alpha$, number of fine-tuning epochs $T$, loss function $\ell$ and learning rate $\eta$.
    \State \textbf{Output:} Fine-tuned model with defense $\mathcal{M}$.
    \State $\mathcal{L}_{init} = \texttt{\MakeUppercase{Calc\_Loss}}(\mathcal{M}_{PT}, \mathcal{D}_f)$
    \State $\mathcal{M} = \texttt{\MakeUppercase{Weights\_Update}}(\mathcal{M}_{PT}, \mathcal{L}_{init}, \eta)$ \Comment{\textbf{Phase \circled{1}}}
    \For {$t$ in $T$}
        \State $\hat{\mathcal{D}}_f^t = \texttt{\MakeUppercase{Data\_Selection}}(\mathcal{M}, \mathcal{D}_f, \mathcal{D}_v)$  \Comment{\textbf{Phase \circled{2}}}
        \State $\mathcal{L}^t = \texttt{\MakeUppercase{Calc\_Loss}}(\mathcal{M}, \hat{\mathcal{D}}_f^t)$
        \State $\mathcal{M} = \texttt{\MakeUppercase{Weights\_Update}}(\mathcal{M}, \mathcal{L}^t, \eta)$  \Comment{\textbf{Phase \circled{4}}}
    \EndFor
    \State \Return $\mathcal{M}$
    \Function{\texttt{\MakeUppercase{Weights\_Update}}}{$\mathcal{M}$, $\mathcal{L}, \eta$}
        \State $\mu = \frac{1}{n} \sum_{i=1}^n \mathcal{L}^i$ \Comment{Average the losses}
        \State $\hat{\theta} = \theta(\mathcal{M}) - \eta \cdot \frac{\partial \mu}{\partial \theta(\mathcal{M})}$ \Comment{Gradients descent}
        \State \Return $\mathcal{M}_{\hat{\theta}}$
    \EndFunction

    \Function{\texttt{\MakeUppercase{Calc\_Loss}}}{$\mathcal{M}$, $\mathcal{D}$}
        \State $\mathcal{L} = \{\}$ \Comment{Initialize the losses}
        \For {$x$ in $\mathcal{D}$}
            \State $l = \ell(\mathcal{M}, x)$ \Comment{Calculate the loss for each sample}
            \State $\mathcal{L} = \mathcal{L} \cup \{l\}$
        \EndFor
        \State \Return $\mathcal{L}$
    \EndFunction

    \Function{\texttt{\MakeUppercase{Data\_Selection}}}{$\mathcal{M}$, $\mathcal{D}_f$, $\mathcal{D}_v$}
        \State $\mathcal{L}_v = \texttt{\MakeUppercase{Calc\_Loss}}(\mathcal{M}, \mathcal{D}_v)$ \Comment{Calculate the validation losses}
        \State $\tau = \frac{1}{n} \sum_{i=1}^n \mathcal{L}_v^i$ \Comment{Use the average validation loss as data selection threshold}
        \State $\hat{\mathcal{D}}_f = \{\}$ \Comment{Initialize the updated fine-tuning dataset}
        \For {$x_f$ in $\mathcal{D}_f$}
            \State $l_f = \ell(\mathcal{M}, x_f)$ \Comment{Calculate the loss for each sample}
            \If {$l_f < \tau$}
                \State $\hat{\mathcal{D}}_f = \hat{\mathcal{D}}_f \cup \{\texttt{\MakeUppercase{Paraphrase}}(x_f, \alpha)$\}  \Comment{\textbf{Phase \circled{3}}}
            \Else
                \State $\hat{\mathcal{D}}_f = \hat{\mathcal{D}}_f \cup \{x_f$\}
            \EndIf
        \EndFor
        \State \Return $\hat{\mathcal{D}}_f$
    \EndFunction
\end{algorithmic}
\end{algorithm}

\subsection{Detailed Algorithm} \label{sec:detail_alg}
Following the overview pipeline, in this section, we offer a detailed description of our approach in Algorithm~\ref{algo:our_tech}.

\noindent
\textbf{Initialization.} (Line 1)
The inputs consist of a pre-trained model, a fine-tuning dataset, a validation dataset, and a set of primary training parameters. Note that $\alpha$ is a scaling factor that determines the strengths for data obfuscation.

\noindent
\textbf{Returning.} (Line 2)
Upon completion of the \Tech{} algorithm, the output is a fine-tuned model with enhanced privacy protection.

\noindent
\textbf{\circled{1} Warm-up Fine-tuning} (Lines 3-4).
This stage utilizes two standard functions: \texttt{\MakeUppercase{Calc\_Loss}} and \texttt{\MakeUppercase{Weights\_Update}}.
\texttt{\MakeUppercase{Calc\_Loss}}, defined in Lines 14-19, computes the loss for each sample in the given dataset $\mathcal{D}$ using the model $\mathcal{M}$. It returns a corresponding list of losses.
\texttt{\MakeUppercase{Weights\_Update}}, defined in Lines 10-13, takes the model $\mathcal{M}$, its loss $\mathcal{L}$, and the learning rate $\eta$, and performs a standard gradient descent step to update the model's weights.
The warm-up stage involves a single pass over the fine-tuning set to help assess the initial influence level of each sample.

\noindent
\textbf{\circled{2} Influential Data Selection} (Lines 6).
During fine-tuning stage, at each epoch $t$, \Tech{} first performs data selection. 
The selection function is defined in Lines 20-30. 
Specifically, \Tech{} calculates the loss for each validation sample and records them in $\mathcal{L}_v$ (Line 21). It then computes the average validation loss in Line 22 as the data selection threshold $\tau$, representing the safe boundary within which MIAs are difficult to succeed. Empirically, the average validation loss is sufficient for the defense.
Subsequently, \Tech{} evaluates each sample from the fine-tuning dataset and select influential ones (Lines 23-29).
If a sample's loss falls below the threshold, i.e., outside the safe range, \Tech{} identifies it as an influential sample. 
Otherwise, the sample is considered safe.
In essence, influential data selection identifies the most influential data points in the fine-tuning set.

\noindent
\textbf{\circled{3} Data Obfuscation} (Line 27).
\Tech{} replaces the identified influential samples with their paraphrased versions. To maintain data quality, we utilize state-of-the-art production LLMs, such as GPT-4~\cite{gpt4} and Claude-3.5~\cite{claude}, employing carefully designed prompts.
The function $\texttt{PARAPHRASE}(x, \alpha)$ in Line 27 takes an original sample $x$ and a paraphrasing strength factor $\alpha$, which controls the extent of paraphrasing applied. For example, setting $\alpha = 0.5$ preserves the first half of the original text while paraphrasing the remaining half.
Tuning $\alpha$ involves a trade-off: a higher paraphrasing ratio may reduce the utility of the data, whereas a lower ratio could increase the risk of privacy leakage. Our ablation study in Section~\ref{sec:exp_ablation} shows that varying $\alpha$ has a slight impact on performance. 
Empirically, we adopt $\alpha = 0.5$ as the default value. 
Further details on paraphrase prompt template are provided in~\cite{SOFT_github}.

\noindent
\textbf{\circled{4} Fine-tuning} (Line 7-8).
Combining the obfuscated data with the remaining safe data, \Tech{} obtains the updated dataset $\hat{\mathcal{D}}_f^t$. The model is then fine-tuned on this updated dataset, resulting in updated model weights.

\Tech{} effectively mitigates membership leakage by replacing influential samples with obfuscated ones. 
With refining the selection of influential data through loss-based prioritization, SOFT balances privacy protection and model utility.

\section{Evaluation}\label{sec:evaluation}
In this section, we provide a comprehensive empirical evaluation of \Tech{} across multiple dimensions.
We describe the experimental setup in Section~\ref{sec:exp_setup}. Section~\ref{sec:exp_main} evaluates the defense effectiveness of \Tech{} against ten MIAs, Section~\ref{sec:exp_utility} quantifies the utility of \Tech{}, and Section~\ref{sec:exp_baseline} demonstrates its superiority over baseline methods. 
Additionally, we conduct adaptive attacks in Section~\ref{sec:adaptive_attack} and a series of ablation studies to investigate the design components in Section~\ref{sec:exp_ablation}.

\subsection{Experimental Setup} \label{sec:exp_setup}

\noindent
\textbf{Datasets.}
Duan et al.~\cite{duan2024membership} were the first to address temporal shift issues in datasets~\cite{min_k_prob}, introducing MIMIR as a novel method for partitioning text samples into training (members) and testing (non-members) sets. 
Following prior works~\cite{duan2024membership, sok_mia_nowhere}, we consider both deduplication strategies and evaluate our approach on six subsets of the Pile dataset~\cite{pile_data}, including ArXiv, HackerNews, PubMed, Pile CC, Wikipedia, and GitHub, with a 13-gram (13\_0.8) data split as the default.

\noindent
\textbf{Models.}
Our analysis primarily focuses on the Llama-3.2~\cite{meta2024llama} model family, including configurations with 1B and 3B parameters.
Unless stated otherwise, the Llama-3.2-3B model is used as the default configuration.
Additionally, we extend our privacy leakage analysis to the Pythia~\cite{pythia} model family, examining various model sizes and datasets in Section~\ref{sec:observations}.

\noindent
\textbf{Attacks Configurations.}
We evaluate our method against 10 different membership inference attacks (MIAs), including both reference-based and reference-free approaches. 
The reference-based attacks include Ratio~\cite{zlib_lowercase_ratio} and Ensemble. The reference-free attacks include Loss~\cite{loss_mia}, Zlib~\cite{zlib_lowercase_ratio}, Lowercase~\cite{zlib_lowercase_ratio}, Min-K\% Prob~\cite{min_k_prob}, Min-K\%++~\cite{min_kpp}, Bag of Words~\cite{sok_mia_nowhere}, ReCall~\cite{recall_mia}, and CON-ReCall~\cite{con_recall}. 
For reference-based attacks, we used OpenLLaMA-7B~\cite{openllama} as the reference model.
In Min-k\% Prob and Min-k\%++, we set $k = 20$. For ReCall and CON-ReCall, we employ a fixed prefix with 10 shots.
Note that some MIAs, i.e., Loss, Zlib, and Lowercase, do not rely on specific hyperparameter choices.

\noindent
\textbf{Defenses Configurations.}
Due to computational constraints, we establish evaluation baselines using DP-LoRA applied to the Llama-3.2-1B model under varying privacy budgets ($\epsilon$).

\begin{table*}[!t]
    \centering
    \fontsize{8}{11}\selectfont %
    \tabcolsep=1.5pt
    \caption{\textbf{Evaluation of \Tech{}'s defense effectiveness against multiple MIAs.} Performance is measured using \textbf{AUC-ROC} scores, where lower values indicate stronger defense.}
    \label{tab:exp_main_auc}
    \begin{tabular}{lcccgggcccgggcccggg}
        \toprule
        \multirow{2.5}{*}{MIAs} & \multicolumn{3}{c}{ArXiv} & \multicolumn{3}{c}{HackerNews} & \multicolumn{3}{c}{PubMed} & \multicolumn{3}{c}{Pile CC} & \multicolumn{3}{c}{Wikipedia} & \multicolumn{3}{c}{GitHub} \\
        \cmidrule(lr){2-4} \cmidrule(lr){5-7} \cmidrule(lr){8-10} \cmidrule(lr){11-13} \cmidrule(lr){14-16} \cmidrule(lr){17-19}
        ~ & Pretrain & FT & \Tech{} & Pretrain & FT & \Tech{} & Pretrain & FT & \Tech{} & Pretrain & FT & \Tech{} & Pretrain & FT & \Tech{} & Pretrain & FT & \Tech{} \\
        \midrule
        Loss~\cite{loss_mia}                    & 0.508 & 0.822 & 0.525 & 0.498 & 0.900 & 0.515 & 0.478 & 0.895 & 0.496 & 0.502 & 0.887 & 0.519 & 0.501 & 0.936 & 0.530 & 0.653 & 0.846 & 0.625 \\
        Zlib~\cite{zlib_lowercase_ratio}        & 0.508 & 0.811 & 0.521 & 0.496 & 0.910 & 0.517 & 0.481 & 0.893 & 0.509 & 0.489 & 0.902 & 0.533 & 0.505 & 0.939 & 0.532 & 0.678 & 0.871 & 0.647 \\
        Lowercase~\cite{zlib_lowercase_ratio}   & 0.490 & 0.785 & 0.517 & 0.507 & 0.845 & 0.515 & 0.515 & 0.850 & 0.541 & 0.482 & 0.858 & 0.522 & 0.499 & 0.887 & 0.536 & 0.611 & 0.820 & 0.591 \\
        Min-K\% Prob~\cite{min_k_prob}          & 0.514 & 0.615 & 0.510 & 0.492 & 0.627 & 0.489 & 0.502 & 0.645 & 0.499 & 0.511 & 0.668 & 0.518 & 0.495 & 0.669 & 0.512 & 0.506 & 0.613 & 0.515 \\
        Min-K\%++~\cite{min_kpp}                & 0.509 & 0.757 & 0.519 & 0.498 & 0.800 & 0.511 & 0.486 & 0.856 & 0.503 & 0.507 & 0.842 & 0.518 & 0.519 & 0.912 & 0.533 & 0.606 & 0.869 & 0.598 \\
        Ratio~\cite{zlib_lowercase_ratio}       & 0.493 & 0.952 & 0.558 & 0.462 & 0.943 & 0.533 & 0.503 & 0.947 & 0.541 & 0.510 & 0.949 & 0.552 & 0.488 & 0.944 & 0.576 & 0.507 & 0.955 & 0.516 \\
        Bag of words~\cite{sok_mia_nowhere}     & 0.504 & 0.508 & 0.505 & 0.529 & 0.521 & 0.523 & 0.513 & 0.528 & 0.518 & 0.483 & 0.504 & 0.511 & 0.501 & 0.507 & 0.507 & 0.701 & 0.649 & 0.660 \\
        ReCall~\cite{recall_mia}                & 0.508 & 0.840 & 0.533 & 0.501 & 0.907 & 0.515 & 0.480 & 0.908 & 0.511 & 0.497 & 0.895 & 0.532 & 0.505 & 0.938 & 0.529 & 0.630 & 0.851 & 0.627 \\
        CON-ReCall~\cite{con_recall}            & 0.505 & 0.764 & 0.518 & 0.486 & 0.740 & 0.500 & 0.488 & 0.868 & 0.516 & 0.458 & 0.844 & 0.513 & 0.496 & 0.925 & 0.530 & 0.638 & 0.847 & 0.620 \\
        Ensemble                                & 0.551 & 0.807 & 0.568 & 0.524 & 0.886 & 0.567 & 0.576 & 0.884 & 0.546 & 0.673 & 0.942 & 0.604 & 0.512 & 0.925 & 0.587 & 0.747 & 0.944 & 0.669 \\
        \midrule
        \textbf{Average}                                 & 0.509 & 0.766 & 0.527 & 0.499 & 0.808 & 0.519 & 0.502 & 0.827 & 0.518 & 0.511 & 0.829 & 0.532 & 0.502 & 0.858 & 0.537 & 0.628 & 0.827 & 0.607 \\
        \bottomrule
    \end{tabular}
\end{table*}

\begin{table*}[t]
    \centering
    \fontsize{8}{11}\selectfont %
    \tabcolsep=1.5pt
    \caption{\textbf{Evaluation of \Tech{}'s defense effectiveness against multiple MIAs.} Performance is measured using \textbf{TPR@1\%FPR} scores, where lower values indicate stronger defense.}
    \label{tab:exp_main_tpr}
    \begin{tabular}{lcccgggcccgggcccggg}
        \toprule
        \multirow{2.5}{*}{MIAs} & \multicolumn{3}{c}{ArXiv} & \multicolumn{3}{c}{HackerNews} & \multicolumn{3}{c}{PubMed} & \multicolumn{3}{c}{Pile CC} & \multicolumn{3}{c}{Wikipedia} & \multicolumn{3}{c}{GitHub} \\
        \cmidrule(lr){2-4} \cmidrule(lr){5-7} \cmidrule(lr){8-10} \cmidrule(lr){11-13} \cmidrule(lr){14-16} \cmidrule(lr){17-19}
        ~ & Pretrain & FT & \Tech{} & Pretrain & FT & \Tech{} & Pretrain & FT & \Tech{} & Pretrain & FT & \Tech{} & Pretrain & FT & \Tech{} & Pretrain & FT & \Tech{} \\
        \midrule
        Loss~\cite{loss_mia}                    & 0.002 & 0.131 & 0.006 & 0.005 & 0.432 & 0.009 & 0.006 & 0.474 & 0.009 & 0.005 & 0.134 & 0.015 & 0.017 & 0.621 & 0.016 & 0.110 & 0.243 & 0.066 \\
        Zlib~\cite{zlib_lowercase_ratio}        & 0.004 & 0.125 & 0.011 & 0.004 & 0.514 & 0.009 & 0.006 & 0.502 & 0.007 & 0.013 & 0.268 & 0.021 & 0.014 & 0.727 & 0.023 & 0.116 & 0.337 & 0.111 \\
        Lowercase~\cite{zlib_lowercase_ratio}   & 0.003 & 0.169 & 0.012 & 0.010 & 0.270 & 0.007 & 0.004 & 0.291 & 0.007 & 0.006 & 0.219 & 0.015 & 0.017 & 0.316 & 0.022 & 0.031 & 0.224 & 0.045 \\
        Min-K\% Prob~\cite{min_k_prob}          & 0.026 & 0.201 & 0.013 & 0.019 & 0.289 & 0.015 & 0.006 & 0.387 & 0.004 & 0.000 & 0.289 & 0.013 & 0.012 & 0.478 & 0.023 & 0.052 & 0.161 & 0.032 \\
        Min-K\%++~\cite{min_kpp}                & 0.009 & 0.072 & 0.007 & 0.013 & 0.195 & 0.012 & 0.007 & 0.385 & 0.008 & 0.002 & 0.152 & 0.014 & 0.009 & 0.598 & 0.023 & 0.056 & 0.301 & 0.055 \\
        Ratio~\cite{zlib_lowercase_ratio}       & 0.005 & 0.892 & 0.021 & 0.005 & 0.700 & 0.020 & 0.007 & 0.765 & 0.037 & 0.004 & 0.896 & 0.093 & 0.014 & 0.884 & 0.057 & 0.028 & 0.891 & 0.051 \\
        Bag of words~\cite{sok_mia_nowhere}     & 0.016 & 0.012 & 0.019 & 0.017 & 0.019 & 0.010 & 0.017 & 0.013 & 0.010 & 0.014 & 0.016 & 0.016 & 0.007 & 0.006 & 0.006 & 0.154 & 0.148 & 0.143 \\
        ReCall~\cite{recall_mia}                & 0.009 & 0.164 & 0.009 & 0.006 & 0.487 & 0.012 & 0.006 & 0.539 & 0.014 & 0.006 & 0.143 & 0.017 & 0.012 & 0.682 & 0.014 & 0.064 & 0.284 & 0.083 \\
        CON-ReCall~\cite{con_recall}            & 0.012 & 0.148 & 0.014 & 0.006 & 0.172 & 0.007 & 0.005 & 0.388 & 0.008 & 0.002 & 0.134 & 0.010 & 0.009 & 0.518 & 0.022 & 0.091 & 0.281 & 0.092 \\
        Ensemble                                & 0.056 & 0.258 & 0.033 & 0.026 & 0.395 & 0.044 & 0.040 & 0.466 & 0.027 & 0.000 & 0.490 & 0.034 & 0.016 & 0.590 & 0.035 & 0.077 & 0.700 & 0.153 \\
        \midrule
        \textbf{Average}                                 & 0.014 & 0.217 & 0.015 & 0.011 & 0.347 & 0.015 & 0.010 & 0.421 & 0.013 & 0.005 & 0.274 & 0.025 & 0.013 & 0.542 & 0.024 & 0.078 & 0.357 & 0.083 \\
        \bottomrule
    \end{tabular}
\end{table*}

\noindent
\textbf{Evaluation Metrics.}
We evaluate both the defense performance and the utility of the fine-tuned model. Defense performance is assessed using the MIA success rate, measured by AUC-ROC and TPR@low\%FPR~\cite{carlini2022first_princeple}. Lower AUC and TPR@low\%FPR values indicate a lower attack success rate and, consequently, higher defense effectiveness.
To evaluate fine-tuned model utility, we use perplexity~\cite{eval_perplexity} and the LLM-as-a-Judge framework~\cite{llm_judge}. Lower perplexity and higher LLM-judge scores indicate greater model utility.
\begin{itemize}[noitemsep, topsep=3pt]
    \item \textbf{AUC-ROC.} The Area Under the Receiver Operating Characteristic Curve (AUC-ROC) measures the performance of a binary classification model by evaluating its ability to distinguish between positive and negative classes across various classification thresholds. 
    Following prior work~\cite{sok_mia_nowhere}, we compute AUC on 1,000 bootstrapped~\cite{bertail2008bootstrapping} subsets of members and non-members, reporting both the mean and standard deviation of the results.
    \item \textbf{TPR@low\%FPR.} This metric, introduced by Carlini et al.~\cite{carlini2022first_princeple}, captures an attack’s ability to confidently identify members of the training set. 
    It is particularly important in high-stakes applications (e.g., medical data or private user information), where even a true positive rate (TPR) around 0.3–0.4 at low false positive rates (FPR) can indicate significant privacy risks. 
    In less sensitive contexts, a TPR@low\%FPR exceeding 0.5 may warrant concern about privacy leakage.
    \item \textbf{Perplexity.} Perplexity reflects the model's confidence in predicting a given sentence. Many previous works rely on perplexity for evaluating LLMs performance on various tasks~\cite{fu2024data,lu2024controlled}.
    We use an open-source evaluation benchmark from EleutherAI~\cite{eval_perplexity} to compute the perplexity score on fine-tuning downstream tasks.
    \item \textbf{LLM-as-a-Judge.}
    Other than perplexity, we adopt the LLM-as-a-Judge framework~\cite{llm_judge} to assess the knowledge learned by the fine-tuned model. Specifically, we utilize a production LLM (e.g., GPT-4o~\cite{gpt4}) to generate multiple QA pairs based on the fine-tuning data and have the fine-tuned model provide answers. The production LLM is further employed to quantitatively evaluate the quality of the model's responses.
\end{itemize}

\subsection{Effectiveness of \Tech{} in Defending MIAs} \label{sec:exp_main}
In this section, we empirically evaluate the effectiveness of \Tech{} against multiple state-of-the-art MIAs (Section~\ref{sec:preliminaries}) and a strong ensemble attack (Section~\ref{sec:observations}). We employ AUC-ROC and TPR@low\%FPR as metrics to measure the attack success rate, where higher values indicate more effective attacks.
Table~\ref{tab:exp_main_auc} presents the AUC-ROC results. The first row lists the six datasets used in our experiments, while the first column enumerates the ten MIAs evaluated.
In this experiment, we utilize Llama-3.2-3B~\cite{meta2024llama} as the default model. For each dataset, we use 1000 samples for fine-tuning and assuming 100 samples for validation purpose.
During the attack phase, the 1000 fine-tuning samples serve as member data and are paired with 1000 distinct samples (not used in fine-tuning) from the same distribution, acting as non-member data. Typically, we fine-tune the model for three epochs and conduct the MIA at the last epoch.
To clearly demonstrate \Tech{}'s effectiveness, we compare its performance with that of the pre-trained model (``Pretrain'') and full fine-tuning (``FT'') for each dataset and MIA, as shown in the second row of the table.
For a fair comparison, the full fine-tuning process employs the same set of samples as \Tech{}. 
Additionally, the last row provides the average scores across all MIAs, offering an overall assessment of performance. 
We also evaluate LoRA fine-tuned models in Table~\ref{tab:lora_auc} and~\ref{tab:lora_tprfpr}.

Notably, for the pre-trained model, the AUC-ROC scores are approximately 0.5 across various MIAs and datasets, except the GitHub dataset.
The fine-tuning dataset is not necessarily disjoint from the pre-training dataset. However, Llama does not disclose details about its pre-training dataset beyond the first version~\cite{meta2024llama}. Even if specific data has been seen during pre-training, successful membership inference attacks on pre-trained LLMs are extremely challenging~\cite{duan2024membership, sok_mia_nowhere}. 
This is not surprising, since each data sample is typically used only once during LLM pre-training~\cite{touvron2023llama, muennighoff2023scaling, komatsuzaki2019one}.
The higher AUC-ROC score for the GitHub dataset compared to others is likely due to the inherent overlap between member and non-member data, as code sharing structural, libraries and function names are more prevalent than natural language~\cite{duan2024membership}. 
In contrast, full fine-tuning is significantly more vulnerable to membership inference attacks, with an average AUC-ROC of 0.819. This vulnerability is consistent with our observations in Section~\ref{sec:observations} \textit{Finding 1 and 2}, which can be exploited by membership inference attacks.
Table~\ref{tab:exp_main_auc} and~\ref{tab:lora_auc} present the AUC-ROC scores, showing that LoRA fine-tuning remains vulnerable to membership inference attacks, especially comparing 
AUC-ROC under attacks such as Ensemble, Ratio, and ReCall.
Comparing with LoRA fine-tuning, \Tech{} effectively reduces attack efficacy by significantly lowering the AUC-ROC scores to 0.540 on average, while LoRA at 0.641. 
SOFT provides significantly better privacy protection, while maintaining much lower perplexity.
Besides, LoRA fine-tuning~\cite{loravsFullfinetune} is highly sensitive to rank choice and parameterization, the model utility can varies under different strategy.
While data obfuscation and selection applied in \Tech{} are more robust.

Table~\ref{tab:exp_main_tpr} and~\ref{tab:lora_tprfpr} present the TPR@1\%FPR results under identical experimental settings. 
Observe that the pre-trained model consistently achieves a TPR@low\%FPR score near 0, indicating minimal vulnerability to attacks. In contrast, full fine-tuning increases the average TPR@1\%FPR to 0.360, LoRA fine-tuning increases average TPR@1\%FPR to 0.181,
demonstrating a higher risk.
However, \Tech{} effectively reduces the average TPR@1\%FPR score to 0.029, closely aligning with the pre-trained model's score and illustrating our defense effectiveness.

Additionally, we observe only a minimal increase in the perplexity score. For example, the perplexity of the pre-trained model on the ArXiv dataset is 12.26. Full fine-tuning reduces this to 9.78, while \Tech{} slightly increases it to 10.49, representing an increase of only 7\%. Despite this modest rise, \Tech{} significantly enhances privacy protection, reducing MIA to near random guessing.

\begin{table}[t]
    \centering
    \fontsize{9}{11}\selectfont %
    \tabcolsep=2.2pt
    \caption{\textbf{Evaluation of LoRA fine-tuning against multiple MIAs.} Performance is measured by \textbf{AUC-ROC} scores.}
    \label{tab:lora_auc}
\begin{tabular}{lcgcgcg}
\toprule
MIAs         & \multicolumn{1}{l}{ArXiv} & \multicolumn{1}{l}{HNews} & \multicolumn{1}{l}{PubMed} & \multicolumn{1}{l}{Pile CC} & \multicolumn{1}{l}{Wiki} & \multicolumn{1}{l}{GitHub} \\
\midrule
Loss         & 0.601                     & 0.645                          & 0.619                      & 0.633                       & 0.644                         & 0.750                      \\
Zlib         & 0.593                     & 0.641                          & 0.621                      & 0.648                       & 0.644                         & 0.776                      \\
Lowercase    & 0.577                     & 0.575                          & 0.595                      & 0.598                       & 0.650                         & 0.716                      \\
Min-K\% Prob & 0.554                     & 0.541                          & 0.550                      & 0.547                       & 0.638                         & 0.643                      \\
Min-K\%++    & 0.584                     & 0.579                          & 0.568                      & 0.549                       & 0.744                         & 0.640                      \\
Ratio        & 0.689                     & 0.702                          & 0.692                      & 0.918                       & 0.774                         & 0.922                      \\
Bag of words & 0.508                     & 0.521                          & 0.528                      & 0.511                       & 0.507                         & 0.651                      \\
ReCall       & 0.582                     & 0.542                          & 0.547                      & 0.545                       & 0.641                         & 0.750                      \\
CON-ReCall   & 0.557                     & 0.577                          & 0.556                      & 0.557                       & 0.627                         & 0.743                      \\
Ensemble     & 0.663                     & 0.749                          & 0.653                      & 0.884                       & 0.847                         & 0.858                     \\
\midrule
Average      & 0.591                     & 0.607                          & 0.593                      & 0.639                       & 0.672                         & 0.745                     \\
\bottomrule
\end{tabular}
\end{table}

\begin{table}[t]
    \centering
    \fontsize{9}{11}\selectfont %
    \tabcolsep=2.2pt
    \caption{\textbf{Evaluation of LoRA fine-tuning against multiple MIAs.} Performance is measured by \textbf{TPR@1\%FPR}.}
    \label{tab:lora_tprfpr}
\begin{tabular}{lcgcgcg}
\toprule
MIAs         & \multicolumn{1}{l}{ArXiv} & \multicolumn{1}{l}{HNews} & \multicolumn{1}{l}{PubMed} & \multicolumn{1}{l}{Pile CC} & \multicolumn{1}{l}{Wiki} & \multicolumn{1}{l}{GitHub} \\
\midrule
Loss         & 0.105                     & 0.209                          & 0.207                      & 0.114                       & 0.116                         & 0.183                      \\
Zlib         & 0.109                     & 0.309                          & 0.313                      & 0.222                       & 0.223                         & 0.645                      \\
Lowercase    & 0.116                     & 0.214                          & 0.209                      & 0.215                       & 0.119                         & 0.156                      \\
Min-K\% Prob & 0.108                     & 0.111                          & 0.115                      & 0.118                       & 0.112                         & 0.112                      \\
Min-K\%++    & 0.106                     & 0.116                          & 0.110                      & 0.114                       & 0.222                         & 0.177                      \\
Ratio        & 0.190                     & 0.161                          & 0.235                      & 0.317                       & 0.337                         & 0.270                      \\
Bag of words & 0.013                     & 0.019                          & 0.013                      & 0.016                       & 0.016                         & 0.141                      \\
ReCall       & 0.105                     & 0.308                          & 0.213                      & 0.117                       & 0.218                         & 0.104                      \\
CON-ReCall   & 0.105                     & 0.205                          & 0.207                      & 0.118                       & 0.216                         & 0.100                      \\
Ensemble     & 0.164                     & 0.371                          & 0.233                      & 0.301                       & 0.374                         & 0.362                        \\
\midrule
Average      & 0.112                     & 0.202                          & 0.186                      & 0.165                       & 0.195                         & 0.225                     \\
\bottomrule
\end{tabular}
\end{table}

\begin{table*}[t]
    \centering
    \fontsize{8}{11}\selectfont %
    \tabcolsep=1.5pt
    \caption{\textbf{Comparison of \Tech{} and DP-LoRA across different noise scales.} Defense effectiveness is evaluated by \textbf{AUC-ROC} against MIAs, where lower values indicate stronger defense. Model utility is measured by perplexity, the lower the better.}
    \label{tab:exp_baseline_auc}
    \begin{tabular}{lgccccccccccg}
        \toprule
        Methods & $\epsilon$ & Loss & Zlib & Lowercase & Min-K\% Prob & Min-K\%++ & Ratio & Bag of words & ReCall & CON-ReCall & Ensemble & Perplexity \\
        \midrule
        Pre-trained               & N/A   & 0.501\textsubscript{±0.014} & 0.493\textsubscript{±0.011} & 0.471\textsubscript{±0.017} & 0.482\textsubscript{±0.013} & 0.509\textsubscript{±0.017} & 0.505\textsubscript{±0.008} & 0.484\textsubscript{±0.010} & 0.483\textsubscript{±0.014} & 0.496\textsubscript{±0.012} & 0.544\textsubscript{±0.009} & 13.19 \\
        \midrule
        \multirow{4}{*}{DP-LoRA}  & 0.01  & 0.504\textsubscript{±0.015} & 0.510\textsubscript{±0.012} & 0.506\textsubscript{±0.008} & 0.504\textsubscript{±0.008} & 0.510\textsubscript{±0.013} & 0.499\textsubscript{±0.011} & 0.515\textsubscript{±0.013} & 0.504\textsubscript{±0.015} & 0.500\textsubscript{±0.010} & 0.553\textsubscript{±0.008} & 13.21 \\
        ~                         & 1     & 0.515\textsubscript{±0.009} & 0.510\textsubscript{±0.008} & 0.503\textsubscript{±0.011} & 0.506\textsubscript{±0.017} & 0.503\textsubscript{±0.011} & 0.497\textsubscript{±0.015} & 0.500\textsubscript{±0.011} & 0.497\textsubscript{±0.007} & 0.493\textsubscript{±0.013} & 0.566\textsubscript{±0.013} & 12.86 \\
        ~                         & 10    & 0.543\textsubscript{±0.013} & 0.549\textsubscript{±0.005} & 0.519\textsubscript{±0.019} & 0.514\textsubscript{±0.013} & 0.523\textsubscript{±0.007} & 0.492\textsubscript{±0.014} & 0.511\textsubscript{±0.014} & 0.520\textsubscript{±0.014} & 0.511\textsubscript{±0.014} & 0.554\textsubscript{±0.006} & 12.65 \\
        ~                         & 20    & 0.596\textsubscript{±0.021} & 0.573\textsubscript{±0.032} & 0.528\textsubscript{±0.009} & 0.524\textsubscript{±0.018} & 0.544\textsubscript{±0.019} & 0.493\textsubscript{±0.026} & 0.501\textsubscript{±0.030} & 0.589\textsubscript{±0.019} & 0.557\textsubscript{±0.013} & 0.684\textsubscript{±0.023} & 12.49 \\
        ~ & 60 & 0.613\textsubscript{±0.013} & 0.605\textsubscript{±0.008} & 0.596\textsubscript{±0.008} & 0.547\textsubscript{±0.011} & 0.567\textsubscript{±0.015} & 0.512\textsubscript{±0.011} & 0.520\textsubscript{±0.008} & 0.629\textsubscript{±0.010} & 0.569\textsubscript{±0.010} & 0.685\textsubscript{±0.009} & 12.36 \\
        ~ & 100 & 0.656\textsubscript{±0.013} & 0.649\textsubscript{±0.012} & 0.631\textsubscript{±0.009} & 0.559\textsubscript{±0.010} & 0.593\textsubscript{±0.007} & 0.552\textsubscript{±0.006} & 0.523\textsubscript{±0.010} & 0.671\textsubscript{±0.010} & 0.583\textsubscript{±0.009} & 0.735\textsubscript{±0.014} & 11.66 \\
        \midrule
        \Tech{}                   & N/A   & 0.517\textsubscript{±0.013} & 0.529\textsubscript{±0.011} & 0.525\textsubscript{±0.011} & 0.506\textsubscript{±0.010} & 0.537\textsubscript{±0.008} & 0.537\textsubscript{±0.008} & 0.517\textsubscript{±0.020} & 0.527\textsubscript{±0.013} & 0.529\textsubscript{±0.012} & 0.573\textsubscript{±0.015} & 11.58 \\
        \bottomrule
    \end{tabular}
\end{table*}

\begin{table*}[t]
    \centering
    \fontsize{8}{11}\selectfont %
    \tabcolsep=1.5pt
    \caption{\textbf{Comparison of \Tech{} and DP-LoRA across different noise scales.} Defense effectiveness is evaluated by \textbf{TPR} \textbf{@1\%FPR}, where lower values indicate stronger defense. Model utility is measured by perplexity, the lower the better.}
    \label{tab:exp_baseline_tpr}
    \begin{tabular}{lgccccccccccg}
        \toprule
        Methods & $\epsilon$ & Loss & Zlib & Lowercase & Min-K\% Prob & Min-K\%++ & Ratio & Bag of words & ReCall & CON-ReCall & Ensemble & Perplexity \\
        \midrule
        Pre-trained               & N/A   & 0.008\textsubscript{±0.004} & 0.008\textsubscript{±0.003} & 0.004\textsubscript{±0.003} & 0.007\textsubscript{±0.002} & 0.020\textsubscript{±0.005} & 0.010\textsubscript{±0.004} & 0.009\textsubscript{±0.002} & 0.006\textsubscript{±0.001} & 0.011\textsubscript{±0.007} & 0.041\textsubscript{±0.010} & 13.19 \\
        \midrule
        \multirow{4}{*}{DP-LoRA}  & 0.01  & 0.001\textsubscript{±0.002} & 0.002\textsubscript{±0.002} & 0.006\textsubscript{±0.003} & 0.014\textsubscript{±0.006} & 0.005\textsubscript{±0.001} & 0.003\textsubscript{±0.003} & 0.005\textsubscript{±0.003} & 0.004\textsubscript{±0.003} & 0.022\textsubscript{±0.004} & 0.020\textsubscript{±0.006} & 13.21 \\
        ~                         & 1     & 0.002\textsubscript{±0.001} & 0.004\textsubscript{±0.003} & 0.009\textsubscript{±0.004} & 0.008\textsubscript{±0.005} & 0.005\textsubscript{±0.002} & 0.002\textsubscript{±0.002} & 0.006\textsubscript{±0.002} & 0.002\textsubscript{±0.002} & 0.008\textsubscript{±0.003} & 0.040\textsubscript{±0.016} & 12.86 \\
        ~                         & 10    & 0.004\textsubscript{±0.006} & 0.004\textsubscript{±0.004} & 0.031\textsubscript{±0.008} & 0.009\textsubscript{±0.005} & 0.000\textsubscript{±0.000} & 0.007\textsubscript{±0.004} & 0.000\textsubscript{±0.001} & 0.012\textsubscript{±0.008} & 0.011\textsubscript{±0.008} & 0.025\textsubscript{±0.011} & 12.65 \\
        ~                         & 20    & 0.017\textsubscript{±0.007} & 0.005\textsubscript{±0.005} & 0.027\textsubscript{±0.018} & 0.009\textsubscript{±0.007} & 0.000\textsubscript{±0.000} & 0.012\textsubscript{±0.008} & 0.015\textsubscript{±0.008} & 0.011\textsubscript{±0.008} & 0.023\textsubscript{±0.015} & 0.021\textsubscript{±0.020} & 12.49 \\
        ~ & 60 & 0.020\textsubscript{±0.005} & 0.010\textsubscript{±0.005} & 0.028\textsubscript{±0.007} & 0.027\textsubscript{±0.010} & 0.008\textsubscript{±0.004} & 0.016\textsubscript{±0.032} & 0.006\textsubscript{±0.003} & 0.010\textsubscript{±0.006} & 0.031\textsubscript{±0.008} & 0.033\textsubscript{±0.018} & 12.36 \\
        ~ & 100 & 0.025\textsubscript{±0.008} & 0.018\textsubscript{±0.009} & 0.027\textsubscript{±0.008} & 0.054\textsubscript{±0.011} & 0.010\textsubscript{±0.005} & 0.055\textsubscript{±0.037} & 0.011\textsubscript{±0.005} & 0.029\textsubscript{±0.006} & 0.033\textsubscript{±0.005} & 0.047\textsubscript{±0.015} & 11.66 \\
        \midrule
        \Tech{}                   & N/A   & 0.003\textsubscript{±0.002} & 0.008\textsubscript{±0.003} & 0.013\textsubscript{±0.003} & 0.016\textsubscript{±0.004} & 0.004\textsubscript{±0.002} & 0.012\textsubscript{±0.005} & 0.008\textsubscript{±0.003} & 0.007\textsubscript{±0.004} & 0.009\textsubscript{±0.006} & 0.032\textsubscript{±0.010} & 11.58 \\
        \bottomrule
    \end{tabular}
\end{table*}

\subsection{Utility Test via LLM-as-a-Judge} \label{sec:exp_utility}
While perplexity is a popular metrics focusing on 
measuring a model's ability to predict word sequences fluently.
To further determine whether the model has truly comprehended the knowledge within its training data~\cite{gao2024train}, we introduce a more comprehensive quantitative assessment of model utility using the LLM-as-a-Judge framework, building on prior work~\cite{llm_judge}.

Human evaluation remains the gold standard for assessing the quality of fine-tuned large language models, however, it is often time-consuming and resource-intensive. To address this challenge, we follow prior work~\cite{llm_judge} implements an automated evaluation approach that utilizes production LLMs, such as GPT-4, as substitutes for human evaluators. This method allows us to efficiently and scalably assess the effectiveness of LLMs fine-tuned with \Tech{}.
Our LLM-as-a-Judge framework operates in two stages. First, it generates questions based on the fine-tuning dataset. Then, it evaluates the responses from the subject model and assigns a corresponding score.

\noindent
\textbf{Template Question Generation.}
We begin by generating 100 evaluation questions using GPT-4o-mini~\cite{gpt4}, prompted with the \texttt{\MakeUppercase{Summarize Prompt}} (detailed prompts in~\cite{SOFT_github}). These questions are designed to assess three key aspects: (1) comprehension of the core technical contributions, (2) understanding of the methodology or approach, and (3) reasoning about the relationships between scientific concepts.

\noindent
\textbf{Qualitative Evaluation.}
We evaluate three models: (1) the pre-trained model, (2) the full fine-tuned model without any defense applied, and (3) the fine-tuned model using \Tech{}. Each model generates responses to the generated questions, which are then scored using GPT-4o-mini via the \texttt{\MakeUppercase{Score Prompt}} (detailed prompts in~\cite{SOFT_github}). This surrogate evaluation assesses the models' performance across three dimensions: content coverage, domain understanding, and response quality. The scores are averaged across all questions to produce a final evaluation score for each model.

\begin{figure}
    \centering
    \includegraphics[width=0.95\linewidth]{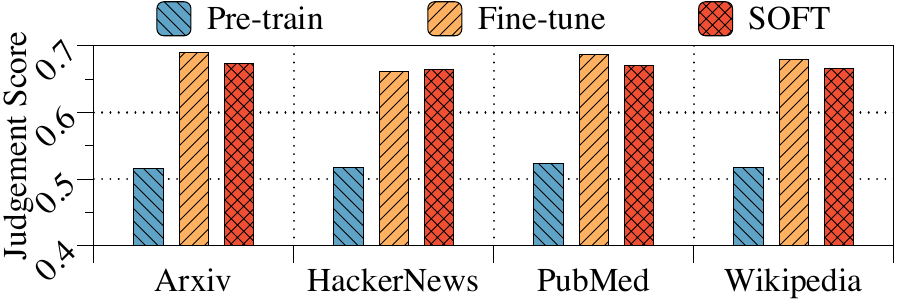}
    \caption{Utility test using LLM-as-a-Judge.}
    \label{fig:QA_llm_as_a_judge}
\end{figure}

Figure~\ref{fig:QA_llm_as_a_judge} presents the results, with the x-axis representing different datasets and the y-axis showing the average judgment score, the higher the better.
Notably, \Tech{} achieves an average score of approximately 0.66, which is only slightly lower than the full fine-tuned model's score of 0.68. This experiment further validates that \Tech{} only introduces minimal loss to the model's utility.

\subsection{Comparison with Defense Baselines} \label{sec:exp_baseline}
We compare the performance of \Tech{} with baseline methods that utilize differential privacy (DP)\cite{dp2006}. DP typically introduces noise to the inputs to prevent the model from overfitting on fine-tuning samples, with the noise scale, $\epsilon$, controlling the strength of DP as defined in Equation~\ref{def:dp_defi}.
Evaluations are conducted using the Llama-3.2-1B model and the ArXiv dataset.
We compare \Tech{} with DP-LoRA across different $\epsilon$ values, evaluating defense performance using AUC-ROC scores from various MIAs and perplexity scores to measure model utility.
As shown in Table~\ref{tab:exp_baseline_auc} and Table~\ref{tab:exp_baseline_tpr}, DP-LoRA struggles to balance the trade-off between defense effectiveness and utility, whereas \Tech{} consistently outperforms DP. Specifically, DP-LoRA with more noise added (e.g. $\epsilon=0.01$) increases the attack efficacy, as indicated by low AUC-ROC scores and TPR, but severely compromises model utility, evidenced by high perplexity scores, approaching those of the pre-trained model.
Conversely, with a high $\epsilon$ (e.g., 20) helps preserve utility to some extent but result in increased privacy leakage.
In contrast, \Tech{} effectively maintains model utility, demonstrated by low perplexity scores, while achieving robust privacy protection against MIAs, demonstrated by low AUC-ROC and TPR@low\%FPR scores. 
When $\epsilon$ at 100, DP-LoRA achieves comparable utility to \Tech{}, however, its privacy protection remains weaker.
We also perform a computational overhead evaluation.
SOFT incurs 15.73\%, while DP-LoRA ($\epsilon$ at 1.0) incur 67.03\%. DP-LoRA need more time to converge, while \Tech{} does not.
These results highlight that \Tech{} successfully balances the trade-off between utility and defense effectiveness, surpassing DP-LoRA in balancing privacy-utility trade-offs.

\subsection{Adaptive Attack}\label{sec:adaptive_attack}
In adaptive attack setting, attackers may modify their strategies to bypass \Tech{}. To measure the resilience of \Tech{} we evaluate three adaptive scenarios.

The first scenario is (1) Paraphrase \& Selection, the adversary knows both paraphrasing model and selection strategy but not its hyperparameters such as temperature or sampling strength. The attacker therefore adopts the same paraphraser model, generates multiple variants of each candidate sample, and aggregates the corresponding membership scores to approximate the original signal.
(2) Paraphrase Only, the attacker possesses full knowledge of the paraphrasing process, including all parameter values, and directly mounts the attack on the paraphrased data. They need independently perform the paraphrasing, due to the generative model, the paraphrased text may still vary.
(3) Selection Only, the paraphraser is disabled. The adversary ignores the paraphrased inputs and attacks only the remaining original un-paraphrased sentences.

Our assessment is carried out on the arXiv dataset, the attacker follows the same methodology of Section~\ref{sec:method} used by the \Tech{} defense algorithm, selecting influential samples and paraphrase them.
As demonstrated in Table~\ref{tab:adaptive}, even when subjected to the adaptive attack, \Tech{}’s performance remains unaffected.
We also provide the ``No Defense (FT)'', which is full fine-tuning model without defense under the Ensemble attack, and ``No Adaptive (w/ \Tech{})'', which applies \Tech{} under the Ensemble attack, for easier comparison.
(1) In the ``Paraphrase \& Selection'' row, 
the attacker remains close to random.
(2) In the ``Paraphrase Only'' row, the attacker shows similar performance with ``Paraphrase \& Selection'' setting. 
(3) In the ``Selection Only'' row, 
this strategy recovers a small number of members with high confidence yet fails on the vast majority, indicating that our data selection strategy is effective, the \Tech{} selected samples are more vulnerable whereas the remaining ones are harder to exploit.

These results demonstrate the robustness of \Tech{} against adaptive attacks.
Distinct large language models produce paraphrases with different styles and distributions that attackers cannot easily reproduce. Furthermore, the dynamic ratio of original and paraphrased data introduced by \Tech{} during training cannot be neutralized by paraphrasing alone.

\begin{table}[!ht]
    \centering
    \tabcolsep=2.2pt
    \caption{Adaptive attack across different settings.}
    \label{tab:adaptive}
\begin{tabular}{lcc}
\toprule
Setting           & \multicolumn{1}{l}{AUC-ROC} & \multicolumn{1}{l}{TPR@1\%FPR} \\
\midrule
No Defense (FT)     & 0.807                 & 0.258 \\
\midrule
Paraphrase \& Selection        & 0.595                       & 0.149                          \\
Paraphrase Only & 0.575                       & 0.136                          \\
Selection Only  & 0.651                       & 0.086                         \\
\midrule
No Adaptive (w/ SOFT)
        & 0.568                             & 0.033 \\
\bottomrule
\end{tabular}
\end{table}

\begin{figure}[ht]
    \centering
    \begin{minipage}[t]{0.5\textwidth}
        \centering
        \includegraphics[width=0.75\textwidth]{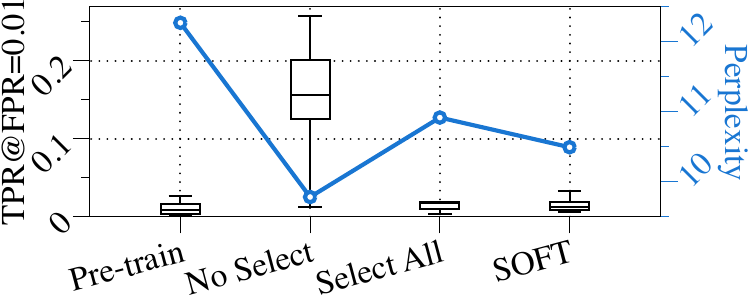}
        \subcaption{ArXiv}
    \end{minipage}
    \begin{minipage}[t]{0.5\textwidth}
        \centering
        \includegraphics[width=0.75\textwidth]{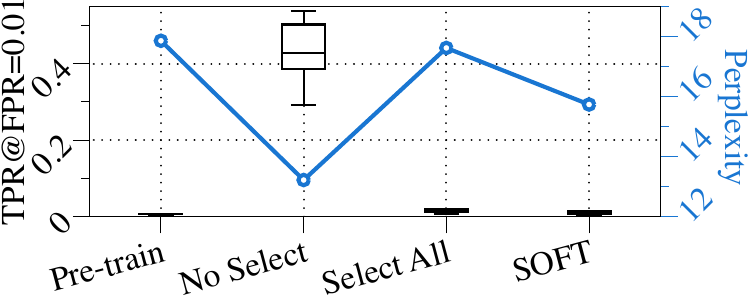}
        \subcaption{PubMed}
    \end{minipage}
    \caption{Ablation study on data selection.}
    \label{fig:ablation_data_select}
\end{figure}

\subsection{Ablation Study} \label{sec:exp_ablation}
In this section, we conduct three ablation studies to assess the impact of \Tech{}'s design component, paraphrasers, and hyper-parameters. Specifically, we study the effect of open-source paraphraser, comparing lexical with semantic, and low-quality paraphraser, data selection, paraphrasing strength $\alpha$ (introduced in Section~\ref{sec:method}), and the integration of \Tech{} with LoRA.

\noindent
\textbf{Effect of Data Selection.}
\Tech{} employs a data selection strategy to balance the trade-off between utility and defense effectiveness. In this study, we investigate the impact of data selection by comparing three approaches: ``No Selection'' (full fine-tuning), ``Full Selection'' (paraphrasing the entire fine-tuning dataset), and our proposed data selection \Tech{}. The results are illustrated in Figure~\ref{fig:ablation_data_select}.
We utilize Llama-3.2-3B as the default model and conduct experiments on the ArXiv and PubMed datasets. 
In both figures, the x-axis represents the design choices, while the left y-axis displays attack efficacy (TPR@1\% FPR), corresponds to the box plot. 
The right y-axis shows the perplexity score on the validation set which indicating model utility, corresponds to the blue lines.
We report the mean score (data point) along with the standard deviation (error bars).
As observed in Figure~\ref{fig:ablation_data_select}, the model without data selection (full fine-tuning) exhibits high TPR@low\%FPR and low perplexity compared to the pre-trained model, indicating enhanced utility but increased vulnerability to MIAs. 
Paraphrasing the entire fine-tuning set (indicated by ``Select All'') reduces the TPR@low\%FPR to nearly 0. However, this approach also degrades utility by increasing the perplexity score. In contrast, \Tech{}'s data selection method reduces the utility loss while achieving comparable levels of privacy protection to full paraphrasing, highlighting the effectiveness of data selection in \Tech{}.

\begin{figure}
    \centering
    \includegraphics[width=0.75\linewidth]{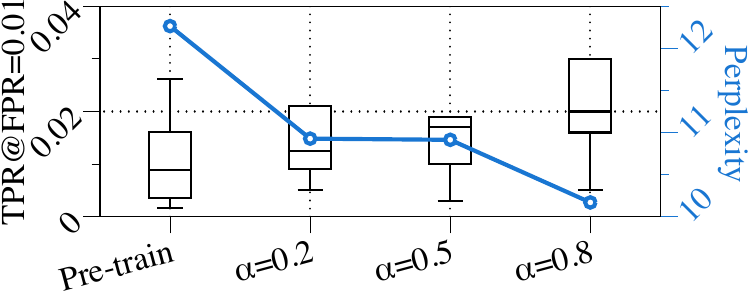}
    \caption{Ablation study on paraphrasing strength $\alpha$.}
    \label{fig:ablation_data_ratio}
\end{figure}

\noindent
\textbf{Effect of Paraphrasers.} We conduct the three paraphrasing ablation studies on arXiv dataset as an example: (1) substituting with open-source model (BART). Other than our evaluation with ChatGPT and Claude in Section~\ref{sec:exp_main}. Here, we include results with BART~\cite{lewis-etal-2020-bart} in Table~\ref{tab:abalation_para}, which offers comparable privacy protection with slightly reduced utility.
(2) Comparing lexical paraphrasing with semantic paraphrasing. We applied semantic paraphrasing in Section~\ref{sec:exp_main}. Here, we include lexical paraphrasing~\cite{ganitkevitch2013ppdb, pavlick2015ppdb} in Table~\ref{tab:abalation_para}, which \Tech{} can still mitigate membership inference risks with a slight utility drop.
(3) Evaluating a lower-quality paraphraser. We evaluate using a small seq2seq model~\cite{sutskever2014sequence, egonmwan2019transformer} for lower-quality paraphrasing in Table~\ref{tab:abalation_para}. We observe that \Tech{} still reduces membership inference risks, with a modest utility trade-off.
\begin{table}[t]
    \centering
    \fontsize{9}{11}\selectfont %
    \tabcolsep=4.5pt
    \caption{Ablation Study on Paraphrasers. T@F denotes TPR@1\%FPR.}
    \label{tab:abalation_para}
\begin{tabular}{lccggcc}
\toprule
\multirow{2}{*}{MIAs} & \multicolumn{2}{c}{BART}                                 & \multicolumn{2}{c}{Lexical}                              & \multicolumn{2}{c}{Seq2Seq}                        \\
    \cmidrule(lr){2-3}\cmidrule(lr){4-5}\cmidrule(lr){6-7}
    ~ & AUC & T@F & AUC & T@F & AUC & T@F \\
\midrule
Loss                  & 0.571                   & 0.137                          & 0.577                   & 0.140                          & 0.616                   & 0.222                          \\
Zlib                  & 0.574                   & 0.136                          & 0.576                   & 0.143                          & 0.616                   & 0.220                          \\
Lowercase             & 0.546                   & 0.122                          & 0.552                   & 0.120                          & 0.619                   & 0.155                          \\
Min-K\%-Prob          & 0.529                   & 0.093                          & 0.525                   & 0.085                          & 0.542                   & 0.114                          \\
Min-K\%++             & 0.561                   & 0.085                          & 0.566                   & 0.088                          & 0.595                   & 0.119                          \\
Ratio                 & 0.563                   & 0.145                          & 0.555                   & 0.147                          & 0.597                   & 0.213                          \\
Bag-of-words          & 0.504                   & 0.010                          & 0.508                   & 0.010                          & 0.508                   & 0.010                          \\
ReCall                & 0.564                   & 0.141                          & 0.566                   & 0.136                          & 0.609                   & 0.212                          \\
CON-ReCall            & 0.514                   & 0.036                          & 0.544                   & 0.051                          & 0.558                   & 0.066                          \\
Ensemble              & 0.555                   & 0.129                          & 0.589                   & 0.095                          & 0.628                   & 0.206                         \\
\bottomrule
\end{tabular}
\end{table}

\noindent
\textbf{Effect of Paraphrasing Strength $\alpha$.}
\Tech{} obfuscates influential data points through proportional paraphrasing, where $\alpha$ denotes the ratio of original text retained. In this experiment, we examine the impact of varying the paraphrasing strength parameter, $\alpha$, by evaluating defense performance at $\alpha$ values of 0.2, 0.5 (default), and 0.8.
We use Llama-3.2-3B as the pre-trained model and fine-tune for 3 epochs by default, and use the ArXiv dataset for these experiments.
The results are depicted in Figure~\ref{fig:ablation_data_ratio}, where the x-axis represents different $\alpha$ values, the left y-axis shows attack efficacy (TPR@1\% FPR), and the right y-axis indicates perplexity scores as a measure of model utility. The outcomes are presented as box plots, displaying the mean and standard deviation.
The outcomes are presented as box plots showing the mean and standard deviation. 
As observed in Figure~\ref{fig:ablation_data_ratio}, larger $\alpha$ values result in improved utility, as indicated by lower perplexity scores (blue lines), but also increase privacy leakage, as reflected by higher TPR@low\%FPR scores (black boxes). The differences in performance metrics across $\alpha$ values are relatively small. Based on these results, we empirically set $\alpha = 0.5$ as the default value for \Tech{}.

\begin{figure}
    \centering
    \includegraphics[width=0.75\linewidth]{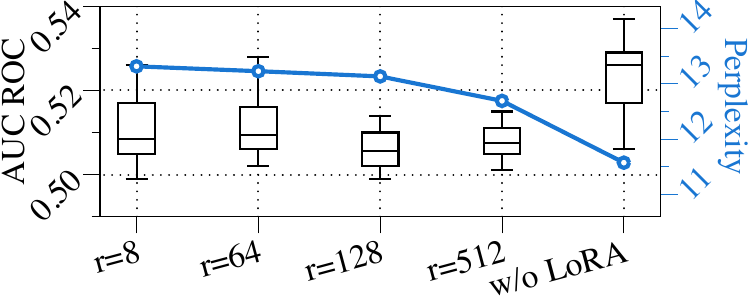}
    \caption{Ablation study on applying LoRA to \Tech{}.}
    \label{fig:ablation_lora}
\end{figure}

\noindent
\textbf{Effect of Applying LoRA in \Tech{}.}
Section~\ref{sec:observations} demonstrates that vanilla LoRA reduces privacy leakage. In this experiment, we evaluate the performance of applying \Tech{} in the context of LoRA fine-tuning.
The results are presented in Figure~\ref{fig:ablation_lora}. The experiment set Llama-3.2-3B as the pre-trained model, and fine-tune via LoRA, and set the ArXiv as dataset.
We investigate the effect of different LoRA rank values, specifically $r$ as defined in Section~\ref{sec:preliminaries}. 
Defense performance is measured using the attack AUC-ROC score, while model utility is assessed using perplexity.
Notably, greater rank reduction (e.g., $r=8$) tends to increase perplexity. 
As expected, as lower ranks typically sacrifices some utility, as reported in the literature~\cite{loravsFullfinetune}. 
However, LoRA further enhances privacy protection, as indicated by the reduced AUC-ROC scores.
This finding aligns with the results from our study on LoRA in Section~\ref{sec:observations} \textit{finding 2}.
Therefore, \Tech{} can be seamlessly integrated with LoRA.

\section{Related Work} \label{sec:related_work}

\noindent
\textbf{LLMs Fine-tuning.}
Researchers fine-tune pre-trained large language models (LLMs) with smaller, task-specific datasets.
To overcome these computational limitations, more efficient fine-tuning methods have been developed,
including Low-Rank Adaptation (LoRA)~\cite{hu2022lora,yin2024lofit,tian2024hydralora},
adapter~\cite{adapter_tuning,hu-etal-2023-llm},
prefix-tuning~\cite{prefix_tuning,zhang2024llamaadapter}, and prompt-tuning~\cite{lester-etal-2021-power,wang2023multitask}.
Among these approaches, LoRA has emerged as a widely adopted method due to its efficiency and effectiveness~\cite{xu2023parameter,zhao2024lora,lora_databricks}. In this paper, we evaluate full fine-tune and LoRA as a representative efficient fine-tuning method.

\noindent
\textbf{LLMs Security.}
LLMs can memorize data~\cite{mireshghallah2022empirical_adapter,carlini2023quantifying,zhang2023counterfactual,su2025mu,guo2024biscope,guoprofiler}, enabling
data extraction attacks to reveal memorized samples~\cite{zlib_lowercase_ratio,yu2023bag,chu-etal-2024-reconstruct,zhang2025censor}.
These security and privacy vulnerabilities~\cite{yan2023parafuzz,shen2024rapid,yan2024aspirer,shen2024bait,guoskewact} are further amplified when LLM-driven autonomous agents~\cite{agents_anthropic} are deployed at scale~\cite{zhang2025llm}. 
To address these risks, various defenses have been proposed, including differential privacy~\cite{dp2006,dp2014, bassily2014private, dpSGD,private_transformer,dpLoRA}, prompt tuning~\cite{dp_prompt_learn}, unlearning~\cite{wang2023kga,warnecke2021machine}, and scrubbing~\cite{pilan2022text}, etc.
However, these defenses often come at the cost of reduced model utility. Building on a systematic analysis, we propose a novel defense method that balances robust privacy protection with high model utility.

\section{Conclusion} \label{sec:conclusion}
We introduce \Tech{}, a novel defense mechanism designed to protect the fine-tuning of large language models against membership inference attacks. Our approach is grounded in the observation that existing MIAs primarily rely on loss or its variants. To mitigate these attacks, \Tech{} selectively replaces influential samples, i.e., those are easily memorized and exhibit lower loss values, with their obfuscated counterparts. 
Our experimental results demonstrate that \Tech{} effectively balances model utility and privacy protection. Specifically, it reduces the attack success rate to near-random guessing while incurring minimal loss in model utility.

\section*{Acknowledgment}
We thank the anonymous reviewers for their constructive comments.
We also thank Chiyuan Zhang for initial discussions during the early stages of this project.
We are grateful to the Center for AI Safety for providing computational resources. This work was funded in part by the National Science Foundation (NSF) Awards CNS-2207204 and IIS-2229876, SHF-1901242, SHF-1910300, Proto-OKN 2333736, IIS-2416835, DARPA VSPELLS - HR001120S0058, ONR N00014-23-1-2081, Amazon and Cisco Research. 
Kaiyuan Zhang is supported in part by the Amazon Fellowship. 
Any opinions, findings and conclusions or recommendations expressed in this material are those of the authors and do not necessarily reflect the views of the sponsors.

\section*{Ethical Considerations}
This paper introduces \Tech{} (\textbf{S}elective data \textbf{O}bfuscation in LLM \textbf{F}ine-\textbf{T}uning), a novel defense mechanism aimed at mitigating privacy risks in fine-tuned large language models. Our work highlights the vulnerabilities of fine-tuned LLMs to membership inference attacks and proposes a practical solution to enhance privacy protection while preserving model utility.
To minimize ethical risks, we used only publicly available datasets from Hugging Face.
The primary stakeholders of this research include individuals whose sensitive data may be used in fine-tuning, researchers and practitioners in the AI and privacy domains, and organizations deploying fine-tuned LLMs in real-world applications.

\section*{Open Science}
In this paper, we replicate the results of various existing membership inference attacks and defenses by using well-established and official GitHub repositories to ensure the validity of our evaluations. To promote transparency and encourage further research on membership inference in LLMs, we release all the attacks and defenses codes, data, and experimental configurations used in our study at 
\href{https://github.com/KaiyuanZh/SOFT}{https://github.com/KaiyuanZh/SOFT}. Artifacts are archived at \href{https://doi.org/10.5281/zenodo.15613620}{https://doi.org/10.5281/zenodo.15613620}.

\bibliographystyle{plain}
\bibliography{reference}

\end{document}